\documentclass[12pt]{article}
\usepackage{fullpage,epsfig,graphics,amsbsy,amssymb,cancel,bbm,cite}
\usepackage{psfrag,hyperref,color,slashed}
\usepackage{graphicx}
\usepackage{amsfonts,mathdots,dsfont}
\usepackage{amssymb,amsmath,amscd,mathrsfs}
\usepackage{tikz}
\usetikzlibrary{arrows,chains,matrix,positioning,scopes,snakes,fadings}

\makeatletter
\tikzset{join/.code=\tikzset{after node path={%
\ifx\tikzchainprevious\pgfutil@empty\else(\tikzchainprevious)%
edge[every join]#1(\tikzchaincurrent)\fi}}}
\makeatother

\tikzset{>=stealth',every on chain/.append style={join},
         every join/.style={->}}
\tikzset{
    >=stealth',
    punkt/.style={
           rectangle,
           rounded corners,
           draw=black, very thick,
           text width=6.5em,
           minimum height=2em,
           text centered},
    pil/.style={
           ->,
           thick,
           shorten <=2pt,
           shorten >=2pt,}
}

\def\alphadot{{\dot{\alpha}}}
\def\betadot{{\dot{\beta}}}

\newcommand{\bea}{\begin{eqnarray}}
\newcommand{\eea}{\end{eqnarray}}
\newcommand{\be}{\begin{equation}}
\newcommand{\ee}{\end{equation}}
\newcommand{\nn}{\nonumber}
\newcommand{\Tr}{\textrm{Tr}}

\def\d{\partial}
\def\bar{\overline}
\def\ha{{1 \over 2}}
\def\b{\bar}
\def\sD{{\scriptscriptstyle D}}

\DeclareMathAlphabet{\mathpzc}{OT1}{pzc}{m}{it}

\usepackage{amsthm}
\usepackage{amsmath}
\usepackage{amsfonts}
\usepackage{amssymb}

\theoremstyle{definition}

\begin{document}
\begin{flushright} \small
UUITP-24/20
 \end{flushright}
\smallskip
\begin{center} \Large
{\bf   S-duality and supersymmetry on curved manifolds}
 \\[12mm] \normalsize
{\bf Guido Festuccia and Maxim Zabzine} \\[8mm]
 {\small\it
   Department of Physics and Astronomy,
     Uppsala University,\\
     Box 516,
     SE-75120 Uppsala,
     Sweden\\
    \vspace{.5cm}
      }
\end{center}
\vspace{7mm}
\begin{abstract}
 We perform a systematic study of S-duality for ${\cal N}=2$ supersymmetric non-linear abelian theories on a curved manifold. Localization can be used to compute certain supersymmetric observables in these  theories. We point out that localization and S-duality acting as a Legendre transform are not compatible.  For these theories  S-duality should be interpreted as Fourier transform and we provide some evidence for this. We also suggest the notion of a coholomological prepotential for an abelian theory that gives the same partition function as a given non-abelian supersymmetric theory. 
\noindent
\end{abstract}

\eject
\normalsize

\tableofcontents
\section{Introduction}

Equivariant localization of quantum field theories on compact manifolds has gained considerable attention since~\cite{Pestun:2007rz} (for a review of the field see~\cite{Pestun:2016zxk}). The localization technique was widely applied to 2D-7D supersymmetric theories on different manifolds that additionally admit some torus action. In this respect two issues need to be addressed: the first problem is to construct supersymmetric field theories on various spaces and to determine which geometrical properties are necessary for supersymmetry. This problem is mainly within classical field theory. The second issue is the implementation of localization for a given supersymmetric problem and involves determining the localization locus and calculating certain superdeterminants.  This appears to be hard in four and higher dimensions where the path integral is often dominated by highly singular configurations that are hard to control on compact manifolds. For example, Pestun's result on $S^4$ \cite{Pestun:2007rz} is largely conjectured by arguing that the path integral is dominated by point-like instantons at the north pole and point-like anti-instantons at the south pole. This should be contrasted with the calculation of the Nekrasov partition function on ${\mathbb C}^2$ \cite{Nekrasov:2002qd, Nekrasov:2003rj} (see also earlier works \cite{Losev:1997tp, Moore:1997dj, Lossev:1997bz, Moore:1998et}) where a well-defined moduli space exists and there is good control (both physical and mathematical) over the singular configurations.  Thus a foremost open problem in localizing on compact manifolds in 4D (and higher dimensions)  is how to define and control the localization locus. In this work, instead of tackling this issue directly, we will approach it from a radically different angle.  
    
This paper is the logical continuation of two our previous works \cite{Festuccia:2018rew} and \cite{Festuccia:2019akm} where we studied ${\cal N}=2$ supersymmetric 4D Yang-Mills on a curved manifold that admits a Killing vector with isolated fixed points. We constructed Killing spinors, defined the corresponding supersymmetry transformations and presented a supersymmetric action. Moreover, by rewriting the theory in appropriate cohomological variables, we showed that it is a generalization of the equivariant Donaldson-Witten theory involving a generalized notion of self-duality for two forms.  The main geometrical data defining the theory is a Killing vector field with isolated fixed points and an assignment of either a plus or minus label to every fixed point. To every neighborhood of a plus fixed point we associate self-dual two forms and to to every neighborhood  of a minus fixed point we associate anti-self-dual two forms. Using the Killing vector field we glue these local bundles in one global sub-bundle of two forms. The localization locus of this theory is controlled by this generalized notion of self-duality and the corresponding PDEs are transversely elliptic.  In  \cite{Festuccia:2019akm} we studied the formal aspects of the transverse ellipticity and its significance for the gauge theory.  At the moment however we do not have a good analytical control of the PDEs that are responsible for the localization locus, hence we can only conjecture the final answer for the partition function of the theory. If the manifold is simply connected we believe that there are two types of contributions to the path integral: point-like instantons attached to plus fixed points (and point-like anti-instantons attached to minus fixed points) and  fluxes which are controlled by $H^2(M, \mathbb{Z})$.  The final conjectured answer for the partition function can be written schematically as follows
 \bea
 \label{coexp}
      Z = \sum\limits_{k_i} \int da ~~e^{2\pi \sum\limits_{i =1}^{p} \frac{1}{\epsilon_i \epsilon_i'}{\cal F}^{\rm ins}_{\rm Nekr} \Big (ia + k_i, \Lambda, \epsilon_i, \epsilon_i' \Big )+ 2\pi \!\!\sum\limits_{i=p+1}^l \frac{1}{\epsilon_i \epsilon_i'}{\cal F}^{\rm anti-ins}_{\rm Nekr} \Big (ia + k_i, \bar{\Lambda}, \epsilon_i, \epsilon_i' \Big ) }~.\label{PF-conj-intro}
\eea  
Here we consider a manifold  $M$ admitting a $T^2$-action and $(\epsilon_i, \epsilon_i')$ characterize the equivariant data for each fixed point $i$. There are $p$ plus fixed points and $(l-p)$ minus fixed points. The overall sum over $k_i$ in front of the integral is due to fluxes controlled by $H^2$. For every fixed point we put a contribution of the Nekrasov partition function 
 $$\frac{\epsilon_1 \epsilon_2}{2\pi} \log Z_{Nekr} = {\cal F}_{Nekr}= {\cal F}^{SW} + O(\epsilon)~.$$ Finally $\Lambda$ (or $\bar\Lambda$) controls the instanton (or anti-instanton) expansion. 
 Versions of  formula (\ref{PF-conj-intro}) have been discussed previously in the context of equivariant Donaldson-Witten theory for  non-compact toric surfaces \cite{MR2227881, Gottsche:2006tn, Gottsche:2006bm, Gasparim:2008ri}, in the compact toric  case \cite{Bawane:2014uka, Sinamuli:2014lma, Rodriguez-Gomez:2014eza, Bershtein:2015xfa, Bershtein:2016mxz}, and in the context of dimensional reduction from 5D to 4D  \cite{Festuccia:2016gul}. 
 
   In this work we take the expression~\eqref{coexp} for granted and study its structural properties. For example, we know that on $\mathbb{R}^4$ the leading term ${\cal F}^{SW}$ is the Seiberg-Witten effective prepotential which enjoys S-duality. So a natural question is how S-duality acts on the answer   \eqref{PF-conj-intro}.  For this we will first study how S-duality acts on an abelian supersymmetric theory (both linear and non-linear) on a curved manifold and investigate the relation between S-duality and supersymmetry. We will then argue that  the localization formula~\eqref{coexp} is compatible with S-duality, provided that this acts by a Fourier transform of each fixed point contribution. The leading term for small $\epsilon$ of the Fourier transform being the Legendre transform one recovers the usual result for ${\cal F}^{SW}$. This is in agreement with earlier studies of the modularity properties of the Nekrasov partition function for arbitrary $\epsilon$'s.  Indeed the relation  between S-duality at finite $\epsilon$'s and the Fourier transform was suggested in~\cite{Galakhov:2012gw}.  
 This idea was further developed in~\cite{Nemkov:2013qma}  based on the explicit study \cite{Billo:2013fi} (also see \cite{Billo:2013jba} for a nice summary of the problem).  
 
 We also suggest that for a non-abelian supersymmetric theory \eqref{PF-conj-intro} one can construct a non-linear $U(1)$ theory that has 
 exactly the same partition function. This theory depends on an effective cohomological prepotential defined as
\bea
\label{copre}
 {\cal F} ({\cal A}, {\mathbf \Lambda}, \chi_{\rm equiv}, \sigma_{\rm equiv} )~,
\eea
where ${\cal A}$ is superfield and all other parameters are replaced by appropriate equivariant cohomology classes including the instanton counting parameter.  The zero form component of ${\mathbf \Lambda}$ approaches $\Lambda$ at some fixed points and $\bar{\Lambda}$ at other fixed points. Here  $\chi_{\rm equiv}$ and $\sigma_{\rm equiv}$ are respectively the Euler and the signature equivariant classes.  The object~\eqref{copre} is related to the Nekrasov partition function and the whole construction is compatible with S-duality acting as the Fourier transform. 
 
The paper is organized as follows: in section \ref{section-2} we review some  facts from \cite{Festuccia:2018rew} about the construction of ${\cal N}=2$ supersymmetric non-abelian gauge theory on a manifold that admits a Killing vector field with isolated fixed points. In particular we stress the use of cohomological variables in the description of the theory. We discuss supersymmetric observables and the Ward identities relating them.   In section \ref{section-3} we concentrate on the abelian version of the theory and explain how S-duality works on a curved manifold.  We argue that S-duality is compatible with supersymmetry. We present arguments both in terms of physical fields and of cohmolological variables. Section \ref{section-4} is devoted to the supersymmetric version of non-linear $U(1)$ theory. Again the main focus is to explain how classical S-duality relates to  supersymmetry. We also comment on the structure of the gravitational corrections and their cohomological description. In section \ref{section-5} we consider the cohomological description of S-duality for the non-linear $U(1)$ theory. We point out that S-duality acting as a Legendre transform is not compatible with the structure of the partition function obtained via localization. As a way to resolve this puzzle we suggest that S-duality should act as a Fourier transform. We consider two examples: $S^4$ and $\mathbb{CP}^2$. In section \ref{section-6} we try to summarize all our discussion and suggest the notion of a cohomological effective prepotential which is supposed to encode the dynamics on the curved manifold. We discuss the possible physical interpretation of this object. Section \ref{section-7} concludes the paper with a summary of the results and a list of open questions, we also briefly comment on the relation between the Fourier transform and the blowup equation for the Nekrasov function. At the end of the paper there are a few appendices with some technical exposition. 

\section{${\cal N}=2$ theory on curved manifolds}\label{section-2}

In this section we review relevant results from \cite{Festuccia:2018rew} and set up the notation we will use.  We focus on the ${\cal N}=2$ vector multiplet with non-abelian gauge group (keeping in mind $U(N)$ as the main example). Nevertheless many considerations can be extended to theories with matter, see \cite{Festuccia:2020yff}.

\subsection{${\cal N}=2$ supersymmetry}\label{subsection-2.1}

A 4D ${\cal N}=2$ vector multiplet contains the gauge field $A$, a complex scalar $X$, an auxiliary real scalar $SU(2)_R$ triplet $D_{ij}$ and gauginos $\lambda_{i\alpha}, \b\lambda_{\dot{\alpha}}^i$ which are in fundamental of $SU(2)_R$. Consider a Riemannian spin manifold $(M,g)$ admitting a smooth real Killing vector $v$ with isolated fixed points and let  $s, \tilde{s}$ be two smooth functions, invariant along $v$ and such that $||v||^2 = s \tilde{s}$.  We can then place on $M$ a ${\cal N}=2$ supersymmetric gauge theory preserving at least one supercharge (see appendices~\ref{n2sugra} and~\ref{app:chivec} for a review of rigid ${\cal N}=2$ supergravity). Note that at each fixed point either $s$ vanishes (we refer to them as $-$ fixed points) or $\tilde{s}$ vanishes (we refer to them as $+$ fixed points).  
Using these geometrical data one can construct Killing spinors and associated supercharges and write a supersymmetric Lagrangian ${\cal L}$:
    \bea
    \label{Lagquadna}
&&(4 \pi) {\cal L} ~{\rm vol}   =  \Tr \Big ( \frac{i}{2} \Big [  \bar{\tau} F^+ \wedge F^+ + \tau F^- \wedge F^- \Big ]  - 2 {\rm Im} (\tau) \Big [ F^+ \wedge W^+ \bar{X} -  F^- \wedge W^- X \Big ]\nn \\
 && +  {\rm Im} (\tau)  \Big [ W^+ \wedge W^+ \bar{X}^2 - W^- \wedge W^- X^2  \Big ]  + \nn \\
 && +  {\rm Im} (\tau) \left[ 4(D^\mu+2 i G^\mu){\bar X}\, (D_\mu-2 i G_\mu)X - \frac{1}{2}  D^{ i j} D_{i j} -  4\Big({R\over 6 }  -N\Big) X{\bar X} \right] {\rm vol} \nn \\
 &&+   {\rm Im} (\tau) \left[ i \lambda_i  \sigma^\mu \Big(\!D_\mu\!+ i G_\mu\Big){\bar \lambda}^{i}+i \bar \lambda^i  \bar\sigma^\mu \Big(\!D_\mu\!-i G_\mu\Big){ \lambda}_{i} \right]{\rm vol} \Big )~.\label{main-action}
   \eea
Here $F$ is the field strength for $A$ and  $\tau$ is defined as 
\bea
 \tau = \frac{\theta}{2\pi} + \frac{4 \pi i}{g^2_{\rm YM}}~. 
\eea
  This Lagrangian also depends on background supergravity fields: the metric $g$, a one form $G$, a two form $W$, a scalar $N$ and a connection for $SU(2)_R$. All these supergravity backgrounds are determined by $v$ and $s, \tilde s$. We refer the readers to \cite{Festuccia:2018rew} for their definitions and detailed expressions. The Lagrangian is invariant (up to boundary terms) under supersymmetric variations that involve the Killing spinors as parameters. The supersymmetry transformations square to a translation along $v$, an $SU(2)_R$-transformation and a gauge transformation.

\subsection{Cohomological description}\label{subsection-2.2}

The formulation of the supersymmetric gauge theory presented in the previous subsection is very obscure from the geometrical point of view. Moreover, the background supergravity fields depend from data whose variation does not change the value of supersymmetric observables. It is instructive to give a cohomological reformulation of the theory which generalizes the equivariant Donaldson-Witten theory. Using the Killing spinors and other geometrical data we can write an invertible map  from the ${\cal N}=2$ vector multiplet $(A, X, \bar{X}, D_{ij}, \lambda_i, \bar{\lambda}^i)$ into another set of fields $(A, \Psi, \phi, \varphi, \eta, \chi, H)$. The new set of fields $(A, \Psi, \phi, \varphi, \eta, \chi, H)$ includes a connection $A$, an odd one-form $\Psi$, two even zero-forms $\phi$  and  $\varphi$, one odd zero form $\eta$, and two forms $\chi$ and $H$ that are odd and even respectively. All these fields except $A$ are in the adjoint of the gauge group. Moreover $P_\omega^+ \chi= \chi$ and $P_\omega^+ H = H$ with $P_\omega^+$ being a generalization of the self-duality projector (see~\eqref{projector} below). The two scalars $(\phi, \varphi)$ are related to the complex scalar $X$ in the vector multiplet as follows
 \bea
\phi = \tilde{s} X + s \bar{X}~,~~~~~~\varphi = - i ( X - \bar{X})~, 
\eea
using which one can borrow the reality conditions from the physical theory.  Note that because the definitions of $\phi$ and $\varphi$ involve both $X$ and $\bar X$ the notions of  holomorphicity in the physical and cohomological variables are not simply related.    In the new cohomological variables the supersymmetry transformations become \footnote{\label{bigfootnote} When discussing the cohomological theory we use the conventions in~\cite{Festuccia:2019akm} that differ from those of~\cite{Festuccia:2018rew}. In particular $\delta^2=\,{\cal L}_v -[\phi-v^\mu A_\mu\,,~\,]$ instead of $\delta^2= i {\cal L}_v -i[\phi+i v^\mu A_\mu\,,~]$. This eliminates factors of $i$ in many formulas. }
     \bea
 && \delta A = \Psi~,\nn \\
 &&  \delta \Psi = \iota_v F + d_A \phi~,\nn\\
 && \delta \phi = \iota_v \Psi~,\nn \\
 && \delta \varphi = \eta~,\label{cohom-transf}\\
 && \delta \eta = {\cal L}^A_v \varphi - [\phi, \varphi]~, \nn \\
 &&\delta \chi = H~,\nn \\
 && \delta H = {\cal L}^A_v \chi - [ \phi, \chi] ~, \nn
 \eea
where $F= dA + A^2$, $d_A  = d + [A, ~]$ and the covariantized Lie derivative ${\cal L}_v^A = d_A \iota_v + \iota_v d_A = {\cal L}_v + [\iota_v A,~ ]$. The transformations square to the Lie derivative and a gauge transformation with parameter $(\phi- \iota_v A)$.  This cohomological field theory formally looks like the equivariant extension of the Donaldson-Witten theory with one important difference, the definition of self-duality on two forms. In the presence of the vector field $v$ it is possible to define a subbundle of $\Omega^2(M)$ of rank 3 that looks in the neighborhood of $+$ fixed points as self-dual two forms and in the neighborhood of $-$ fixed points as anti-self dual two forms. This subbundle can be defined by means 
of the following projector
    \bea
    \label{projector}
   P_\omega^+ = \frac{1}{1+ \cos^2 \omega} \Big ( 1 + \cos \omega \star - \sin^2 \omega  \frac{\kappa \wedge \iota_v}{\iota_v \kappa} \Big )~,
  \eea
  where the one form $\kappa=g(v)$ and 
   \bea
    \cos \omega = \frac{s - \tilde{s}}{s+ \tilde{s}} ~.
   \eea
 This projector is well-defined at the fixed points and is naturally related to supersymmetry on $M$.  There are alternative ways to describe this projector and the corresponding subbundle of two forms, see \cite{Festuccia:2018rew} for further details. The case when all fixed points are plus (or minus) corresponds to the equivariant Donaldson-Witten theory. 
 
 If we want to localize, we need to add additional fields to deal with the gauge symmetry, $(c, \bar{c}, b)$: ghost, anti-ghost and a Lagrangian multiplier. The resulting cohomological theory is controlled by a transversely elliptic complex (see \cite{Festuccia:2019akm}) and the corresponding  localization locus is described in terms of transversely elliptic PDEs (or various deformations thereof). It is hard to say something definite about the localization locus beside a general conjecture that the path integral is dominated by point like (anti)-instantons and fluxes when the manifold is simply connected. 
     
   Leaving aside the complications related to the details of the localization locus we can make many observations on general grounds. One key comment is that the action \eqref{main-action} rewritten in cohomological variables has the following structure
    \bea
     \int\limits_{M}   4\pi {\cal L} \, {\rm vol} = \int\limits_{M}  \Tr (\phi + \Psi + F )^2 (\Omega_0 + \Omega_2 + \Omega_4)  + \delta (...)~,\label{cohomol-action}
    \eea
    up to BRST-exact terms. The multi-form $\Omega=(\Omega_0 + \Omega_2 + \Omega_4)$ is defined as follows
  \bea
 && \Omega_0 = \frac{\tau s + \bar{\tau} \tilde{s}}{s+ \tilde{s}}~,\nn \\
 && \Omega_2 = - (\tau - \bar{\tau}) \frac{s-\tilde{s}}{(s+\tilde{s})^3} d\kappa - 2 \frac{(\tau - \bar{\tau})}{(s+ \tilde{s})^3} \kappa \wedge d(s-\tilde{s})~,\label{cohomol-observ-ex}\\
 && \Omega_4 = 3(\tau - \bar{\tau}) \frac{s-\tilde{s}}{(s+\tilde{s})^5} d\kappa \wedge d \kappa + 12 \frac{(\tau - \bar{\tau})}{(s+\tilde{s})^5} \kappa \wedge d\kappa \wedge d(s-\tilde{s})~,\nn
 \eea
 and it is closed under $d_v = d + \iota_v$. Up to  BRST-exact terms the action depends only on the class $\Omega = \Omega_0 + \Omega_2 + \Omega_4$ in $H_{\rm equiv}(M)$. In the next subsection we discuss the formal consequences of this observation. 

\subsection{Ward identities and localization}\label{subsection-2.3}

In this subsection we would like to explore general aspects of localization and Ward identities.  This discussion is formal and it is applicable to a wide class of theories in different dimensions. It allows us to discuss the general features of equivariant theories independently from the concrete PDEs which describe their localization locus. 
      
Following the terminology from \cite{Losev:1997tp} we concentrate on the ``holomorphic'' part of the cohomological theory \eqref{cohom-transf} which is defined by the following transformations
 \bea
&&  \delta A = \Psi~,\nn \\
&&  \delta \Psi = \iota_v F + d_A \phi~,\\
&&  \delta \phi = \iota_v \Psi~. \nn
 \eea
 It is natural to combine these transformations as
 \bea
  \delta \Big (\phi + \Psi + F \Big ) = (d_A + \iota_v) \Big (\phi + \Psi + F \Big )~,\label{expicit-hol-mult}
 \eea
 where $F$ is the field strength for $A$. 
 If we take any invariant polynomial $P$ on the corresponding Lie algebra  then 
   \bea
  \delta~ P [\phi + \Psi + F  ] = (d + \iota_v) ~P [\phi + \Psi + F ]~,
 \eea
 which we can multiply by any equivariantly closed form $\Omega = \Omega_0 + \Omega_2 + \Omega_4$ 
   \bea
    d_v \Big ( \Omega_0 + \Omega_2 + \Omega_4 \Big )=0~,
    \eea
 where $d_v = d + \iota_v$. 
  As result we construct a collection of differential forms that satisfy
    \bea
    (\delta -  d_v) \Big ( P [\phi + \Psi + F  ] \Omega \Big )=0 ~. \label{relation-diffs}
   \eea
    Let us  consider a concrete choice of invariant polynomial and define
   \bea
    \Tr (\phi + \Psi + F )^2 (\Omega_0 + \Omega_2 + \Omega_4) = \omega_0 + \omega_1 + \omega_2 + \omega_3 + \omega_4 = \boldsymbol{\omega}(x) ~,\label{defin-observ-forms}
   \eea
   where the forms $\omega_i$ are 
   \bea
   && \omega_0 = \Tr(\phi^2) \Omega_0~, \nn \\
   && \omega_1 = \Tr (2 \Psi \phi) \Omega_0~, \nn\\
   && \omega_2 = \Tr (\phi^2) \Omega_2 + \Tr (2\phi F + \Psi^2) \Omega_0~,\\
   && \omega_3 = \Tr (2\Psi \phi) \Omega_2 + \Tr (2 \Psi F) \Omega_0~, \nn \\
   && \omega_4 = \Tr (F^2) \Omega_0 + \Tr (2\phi F + \Psi^2) \Omega_2 + \Tr (\phi^2) \Omega_4~. \nn
   \eea
    The symmetry property \eqref{relation-diffs} implies the following descent relations
   \bea
    && \delta \omega_0 = \iota_v \omega_1~,\\
    && \delta \omega_1 = d \omega_0 + \iota_v \omega_2~,\\
    && \delta \omega_2 = d \omega_1 + \iota_v \omega_3~,\\
    && \delta \omega_3 = d \omega_2 + \iota_v \omega_4~,\\
    && \delta \omega_4 = d \omega_3~. 
   \eea
    If we are interested in observables (objects annihilated by the BRST differential $\delta$) then 
   the only local observable is $\omega_0(x_i)$  where $x_i$ is one of the fixed points. 
    The observable 
    \bea
     \int\limits_{\gamma} \omega_1
    \eea
    is supersymmetric if the one-cycle $\gamma$ is invariant under our action (i.e. it is along $v$) and so on. Hence we construct  observables as integrals of $\omega_i$ over invariant $i$-cycles. 
     The top observable is given by
     \bea
      \int\limits_{M} \Big (  \Tr (F^2) \Omega_0 + \Tr (2\phi F + \Psi^2) \Omega_2 + \Tr (\phi^2) \Omega_4 \Big )~,\label{observ-int}
     \eea
      and this is exactly the observable which appears in \eqref{cohomol-action} for a specific choice of $\Omega$. 
   All observables depend only on the cohomology class of $ ( \Omega_0 + \Omega_2 + \Omega_4)$ within $H_{\rm equiv}(M)$. 
   
   The next question to ask is if there are any non-trivial relations between the expectation values of these observable, any Ward identities that relate them. We assume the following property of the path integral 
    \bea
     \int \delta \Big ( ... e^{S} \Big ) = 0~,
    \eea
    where $S$ is some $\delta$ invariant action. Now consider the collection of forms $\boldsymbol{\omega}(x)$ defined in \eqref{defin-observ-forms} satisfying $(\delta - d^x_v) \boldsymbol{\omega}(x)=0$ (here the upper script $x$ indicates on which variable $d_v$ acts).   
      Thus we get the following collection of Ward identities
      \bea
     &&  d_v^x \langle \boldsymbol{\omega}(x) \rangle = 0~,\nn\\
     && (d_v^{x} + d_v^{y})  \langle \boldsymbol{\omega}(x) \boldsymbol{\omega}(y) \rangle =0 ~, \\
     && (d_v^{x} + d_v^{y} + d_v^z)  \langle \boldsymbol{\omega}(x) \boldsymbol{\omega}(y) \boldsymbol{\omega}(z) \rangle =0~,\qquad {\rm  etc.}\nn
      \eea
 Here we understand the correlator  $\langle \boldsymbol{\omega}(x) \rangle$ as an element of $\Omega^\bullet(M)$, the correlator $\langle \boldsymbol{\omega}(x) \boldsymbol{\omega}(y) \rangle$ as an element of $\Omega^\bullet(M\times M)$ etc. In these Ward identities the equivariant differential is defined with respect to a diagonal action on the factors. For example if $v$ corresponds to a $T^2$-action on $M$ then the differential $(d_v^{x} + d_v^{y})$ corresponds to $T^2$ action on $M\times M$ ($T^2$ acts identically on two factors).  Assuming that we deal with equivariantly closed smooth differential forms we apply the localization theorem and get
  \bea
&&  \int\limits_M  \langle \omega_4(x) \rangle = 2\pi \sum\limits_{i} \frac{1}{\epsilon_i \epsilon'_i}  \langle \omega_0 (x_i) \rangle \\
 &&  \int\limits_{M\times M}   \langle \omega_4(x)\omega_4 (y)  \rangle = (2\pi)^2 \sum\limits_{i,j} \frac{1}{\epsilon_i \epsilon'_i \epsilon_j \epsilon'_j}  \langle \omega_0 (x_i)  \omega_0 (x_j) \rangle\qquad {\rm  etc. }  \label{two-point-loc}
 \eea
Here $x_i$ are fixed points on $M$ and $(\epsilon_i, \epsilon_i')$ can be read off from the local action of $T^2$ at $x_i$.
 Thus we can expect that 
 \bea
  \Big \langle ~e^{\int\limits_M \omega_4}~ \Big \rangle = \Big \langle~ e^{2\pi \sum\limits_{i} \frac{1}{\epsilon_i \epsilon'_i}  
  \omega_0 (x_i)} ~\Big \rangle~.
 \eea          
  These formal Ward identities lead to localization, although they do not help to carry out concrete calculations.
  
In the logic presented above there is a loophole, namely the assumption that that the correllator  $\langle \boldsymbol{\omega}(x) \boldsymbol{\omega}(y) \rangle$ is a smooth differential form on $M\times M$.  Actually it is more natural to expect that this correlator  is a distribution on $M\times M$ with some $\delta$-function like behaviour on the diagonal $x=y$. We would expect that the correlator $\langle \boldsymbol{\omega}(x) \boldsymbol{\omega}(y) \rangle$ is a smooth differential form away from the diagonal. However removing the diagonal from $M \times M$ makes the space non-compact and one cannot apply the localization argument directly to this non-compact space. There are two possible scenarios and which one is realized may depend on the details of the theory. The first possibility is that the singularity on the diagonal is rather mild and the left hand side of \eqref{two-point-loc} (the integration of the top form of the two point correlator) is well-defined without any additional contact terms. In this situation we would expect that localization still works and the result \eqref{two-point-loc} holds true. The second scenario involves the analysis of possible contact terms on the diagonal. However if we require that  supersymmetry is preserved, then the contact term on the diagonal should be supersymmetric by itself  and thus can be localized on the diagonal by itself. Let us illustrate this schematically. Consider the combination 
  \bea
    \langle \boldsymbol{\omega}(x) \boldsymbol{\omega}(y) \rangle  -   \langle \boldsymbol{\omega}^2(x) \rangle G(x-y) ~, 
    \label{combin-contact-term}
  \eea
 where we have assumed that the contact term has this structure with $G(x-y)$ being a top form concentrated on the diagonal (some sort of $\delta$-function) with the property $\int dy  ~G(x-y) =1$.  If we assume that the combination \eqref{combin-contact-term} is a smooth  form on $M\times M$ then we can apply localization and the result will look as follows
\bea
 \int\limits_{M\times M}   \langle \omega_4(x)\omega_4 (y)  \rangle  = (2\pi)^2 \sum\limits_{i,j} \frac{1}{\epsilon_i \epsilon'_i \epsilon_j \epsilon'_j}  \langle \omega_0 (x_i)  \omega_0 (x_j) \rangle   + 2\pi  \sum\limits_{i} \frac{1}{\epsilon_i \epsilon'_i}  \langle \omega^2_0 (x_i) \rangle~,
 \label{double-local}
\eea
 where we  localized both on $M\times M$ and on $M$ for the second term in \eqref{combin-contact-term}.  The present ansatz \eqref{combin-contact-term} is ad hoc but the important property is that the contact term is supersymmetric on its own and thus can be localized. The concrete details of possible contact terms may depend on the theory, however, if we assume that they are supersymmetric then we should always obtain some version of formula \eqref{double-local}. 
  
  The present discussion of Ward identities for an equivariant cohomological theory is formal. The main lesson is that unlike in the standard cohomological  theory in the equivariant theory there are additional relations between different observables. 

\section{S-duality for abelian ${\cal N}=2$ theory}\label{section-3}

In this section we consider the abelian version of ${\cal N}=2$ theory described in the previous section. We show that the coupling to rigid supergravity is consistent with S-duality.  We also introduce some concepts and technical tools that will be used in the following sections. 

\subsection{{${\cal N}=2$  supersymmetric theory}}\label{subsection-3.1}

We will start by recasting the abelian version of the Lagrangian~\eqref{Lagquadna} in a way suitable to study S-duality. For this we introduce a chiral multiplet $X$ of weight $w=1$ and an anti-chiral multiplet $\bar X$ also of weight $w=1$. Here and in the following we will use the same letter (e.g. X) to identify both a chiral multiplet and its lowest component. The component expansion and properties of ${\cal N}=2 $ chiral multiplets and vector multiplets in a rigid supergravity background are reviewed in appendix \ref{app:chivec}.

 We consider the supersymmetric quadratic Lagrangian:
\bea\label{lagdualb}
{\cal L} =&-{i\over 4\pi}\left[\bar \tau \left(T(X^2)+{1\over 2} W^-_{\mu\nu}W^{-\mu\nu} X^2\right) -\tau \left( \b T(\b X^2)+{1\over 2} W^+_{\mu\nu}W^{+\mu\nu} \b X^2\right)\right]~,
\eea
where $T(X^2)$ is the top component of the chiral multiplet whose lowest component is $X^2$ (see \eqref{multrules}). We also introduce a vector multiplet whose components will be denoted via the subscript $D$ and add the following couplings:
\bea\label{lagdualbc}
-{i\over 2\pi}\left[T( X X_{\scriptscriptstyle D})+{1\over 2} W^-_{\mu\nu}W^{-\mu\nu} X X_{\scriptscriptstyle D}-\b T (\b X \b X_{\scriptscriptstyle D})-{1\over 2} W^+_{\mu\nu}W^{+\mu\nu} \b X \b X_{\scriptscriptstyle D} \right]~.
\eea
Using the product rules~\eqref{multrules} we can expand the resulting Lagrangian and obtain:
\bea\label{lagrpartsb}
{\cal L}=&&{i\over 8\pi}(\bar \tau B^{+\mu\nu}B^+_{\mu\nu}-\tau B^{-\mu\nu}B^-_{\mu\nu})+{i\over 4\pi}\epsilon^{\mu\nu\rho\lambda}\left(B_{\mu\nu}+ \bar X W^+_{\mu\nu}+ X W^-_{\mu\nu}\right)\partial_\rho A_{{\scriptscriptstyle D}\lambda}+\cr
&&-{i\over 2\pi}\left[\bar \tau \left(T X+{1\over 2} W^-_{\mu\nu}W^{-\mu\nu} X^2\right)-\tau \left(\b T \b X+{1\over 2} W^+_{\mu\nu}W^{+\mu\nu} \b X^2\right)\right]+\cr
&&-{i\over 2\pi}\left(X_{\scriptscriptstyle D} T+X  (D^\mu+2i G^\mu)(\partial_\mu+2i G_\mu)\b X_{\scriptscriptstyle D}-\bar X_{\scriptscriptstyle D} \bar T -\bar X (D^\mu-2i G^\mu)(\partial_\mu-2i G_\mu) X_{\scriptscriptstyle D}\right)+\cr
&&-{i\over 2\pi}\left({1\over 6} R-N\right)(X \b X_{\scriptscriptstyle D}-\b X X_{\scriptscriptstyle D})-{i\over 4\pi}(B^+_{\mu \nu} W^{+\mu\nu} \b X_{\scriptscriptstyle D} -B^-_{\mu\nu} W^{-\mu\nu} X_{\scriptscriptstyle D})~+\cr
&& -{i\over 16\pi}(\bar \tau D_{i j} D^{i j} -\tau \b D_{i j} \b D^{i j })-{i\over 8\pi}(D_{i j} D_{\scriptscriptstyle D}^{i j}-\b D_{i j} D_{\scriptscriptstyle D}^{i j}) +
{1\over 4\pi}(\bar \tau \lambda_i \psi^i-\tau \b \lambda^i\b \psi_i)+\cr &&+{1\over 4\pi}\left(\lambda_{{\scriptscriptstyle D} i} \psi^i-\lambda_i  \sigma^\mu (D_\mu+i G_\mu) \b \lambda_{\scriptscriptstyle D}^i-\b\lambda_{\scriptscriptstyle D}^i \b \psi_i + \b \lambda^i \b \sigma^\mu(D_\mu-i G_\mu) \lambda_{{\scriptscriptstyle D} i}\right)~.
\eea
The vector multiplet components $X_{\scriptscriptstyle D},\bar X_{\scriptscriptstyle D}, A_{\scriptscriptstyle D}, \lambda_{\scriptscriptstyle D},\b \lambda_{\scriptscriptstyle D}, D^{i j}_{\scriptscriptstyle D}$ appear linearly and can be integrated out. This enforces constraints on the $X$ and $\bar X$ multiplets that get shortened to a vector multiplet according to~\eqref{vectorconst}. The final result is the abelian version of the  Lagrangian~\eqref{Lagquadna} with coupling constant $\tau$. 

Alternatively we can integrate $T,\bar T, \psi,\bar \psi, B^+, B^- ,D,\b D$ that also appear linearly to obtain:
\bea\label{intdualb}
&&X=-{1\over \bar \tau}X_{\scriptscriptstyle D}~,\quad \bar X= -{1\over \tau} \b X_{\scriptscriptstyle D}~,\cr
&&\lambda^i=-{1\over \bar \tau } \lambda_{\scriptscriptstyle D}^i~, \quad \bar \lambda^i =-{1\over \tau} \b\lambda^i_{\scriptscriptstyle D}~,\cr
&& B^+=-{1\over \b \tau} \left(dA_{\scriptscriptstyle D}^+ - W^+ \b X_{\scriptscriptstyle D}\right)~,\quad B^-=-{1\over \tau} \left(dA_{\scriptscriptstyle D}^- -W^- X_{\scriptscriptstyle D}\right)~,\cr
&& D^{i j}=-{1\over \b\tau}D_{\scriptscriptstyle D}^{i j}~,\quad \b D^{i j}=-{1\over \tau} D_{\scriptscriptstyle D}^{i j}~.
\eea
Plugging back  we get a Lagrangian for the vector multiplet. Again this will be as in~\eqref{Lagquadna} but now with coupling constant $-{1\over \tau}$. Hence we see that S-duality is compatible with coupling to a rigid supergravity background at least in the case of a free abelian vector multiplet.

\subsection{S-duality in cohomological variables}\label{subsection-3.2}

Here we want to reformulate the discussion from the previous subsection in terms of cohomological field theory.  In our treatment we follow closely ideas from \cite{Losev:1997tp}  which we generalize to the case of equivariant cohomological field theory. 

 As we have reviewed the ${\cal N}=2$ vector multiplet can be mapped to the cohomological variables   $(A, \Psi, \phi, \varphi, \eta, \chi, H)$ described in subsection \ref{subsection-2.2}.  Following the terminology from  \cite{Losev:1997tp} we can refer to $(A, \Psi, \phi)$ as a holomorphic multiplet and the rest of the fields as non-holomorphic variables. The holomorphic multiplet combines naturally in a superfield (short superfield)
   \bea
{\cal A}_\sD = \phi_\sD + \Psi_\sD + F_\sD~,\label{short-multiplet}
 \eea
  where we put the subscript $D$.  This multiplet transforms as follows 
  \bea
   \delta {\cal A}_\sD = (d + \iota_v) {\cal A}_\sD~,
   \eea 
   see equation \eqref{expicit-hol-mult}. If we consider the cohomological description of the chiral multiplet (see Appendix \ref{app:chiral}) we can analogously split the multiplet into an holomorphic part and a non-holomorphic part. The holomorphic part can be combined in a long multiplet
 \bea
 {\cal A} = \phi + \Psi + F + \rho + D~,\label{long-multiplet}
 \eea
  where we have forms of all degrees of alternating parity ($\Psi$ and $\rho$ are respectively fermionic one and three forms) and $F$ is now an arbitrary two form. The supersymmetry acts as follows on the long superfield
   \bea
     \delta {\cal A} = (d+\iota_v) {\cal A}~.
  \eea
    We can write the following supersymmetric action
   \bea
     S= \int i {\cal A} {\cal A}_\sD + \frac{i}{2} \Omega {\cal A}^2=
      \int i {\cal A} {\cal A}_\sD + \frac{i}{2} (\Omega_0 + \Omega_2 + \Omega_4) {\cal A}^2 \label{quadratic-action}
   \eea
  provided that the collection of background forms $\Omega$ satisfies
  \bea
   (d + \iota_v) (\Omega_0 + \Omega_2 + \Omega_4) =0~. 
  \eea
   Actually if we shift $\Omega$ by $d_v \alpha$ (assuming that ${\cal L}_v \alpha=0$) the action \eqref{quadratic-action} changes by a $\delta$-exact term.  Thus cohomologically the action  \eqref{quadratic-action} depends only on the class of $\Omega$ in $H_{\rm equiv}(M)$. The action  \eqref{quadratic-action} has the following expansion in components
     \bea
      S=  i \int \Big [ F F_\sD + \rho \Psi_\sD + D \phi_\sD + \frac{1}{2} \Omega_4 \phi^2 + \Omega_2 \Big ( \phi F + \frac{1}{2} \Psi^2 \Big ) + \Omega_0 \Big (\phi D + \Psi \rho + \frac{1}{2} F^2 \Big ) \Big ] ~.    
     \eea
  Integrating out $\phi_\sD$, $\Psi_\sD$ and $F_\sD$ the long multiplet collapses to the short one. The integration over $\phi_\sD$ sets $D=0$, the integration over $\Psi_\sD$ sets $\rho=0$ and the integration over $F_\sD$ impose the constraint that $F$ is the curvature of a line bundle. As usual (see e.g.\cite{Witten:1995gf}) the integration over $F_\sD$ combines a sum over line bundles and an actual integration.  After these integrations we arrive at the following action 
    \bea
    S= \frac{i}{2} \int (\Omega_0 + \Omega_2 + \Omega_4)  (\phi + \Psi + F)^2~,
   \eea
   where now only the holomorphic part of the vector multiplet appears. This is an example of the observable discussed in the previous section.  
   
 Alternatively if we integrate out $F$, $\rho$ and $D$ in the action  \eqref{quadratic-action} we obtain the following relations
 \bea
  && \phi = - \frac{1}{\Omega_0} \phi_\sD~,\\
  && \Psi = - \frac{1}{\Omega_0} \Psi_\sD~,\\
  && F = - \frac{1}{\Omega_0} F_\sD - \frac{\Omega_2}{\Omega_0} \phi = - \frac{1}{\Omega_0} F_\sD + \frac{\Omega_2}{\Omega^2_0} \phi_\sD~, 
 \eea
 and as result we have 
 \bea
  \iota_v F + d\phi = - \frac{1}{\Omega_0} \Big (\iota_v F_\sD + d\phi_\sD \Big )~. 
 \eea
  Let us assume for the moment that $\Omega_0^{-1}$ is well-defined. 
  If we evaluate the action we get 
    \bea
    S= \frac{i}{2} \int \frac{-1}{(\Omega_0 + \Omega_2 + \Omega_4)}  (\phi_\sD + \Psi_\sD + F_\sD)^2~,\label{dual-observ-2}
   \eea
    where the inverse $\Omega^{-1}$ is understood as follows
    \bea
     \frac{-1}{(\Omega_0 + \Omega_2 + \Omega_4)} = - \frac{1}{\Omega_0} + \frac{\Omega_2}{\Omega_0^2} + \frac{\Omega_4}{\Omega_0^2} - \frac{\Omega_2^2}{\Omega_0^3}~.\label{inverse-forms}
    \eea
  One can check explicitly that if $d_v \Omega=0$ then 
  \bea
   d_v \left (  - \frac{1}{\Omega_0} + \frac{\Omega_2}{\Omega_0^2} + \frac{\Omega_4}{\Omega_0^2} - \frac{\Omega_2^2}{\Omega_0^3} \right ) =0~. 
  \eea
  Thus the expression \eqref{dual-observ-2} is a supersymmetric observable in the dual theory.  Next we have to argue that under S-duality the concrete representative for $H_{\rm equiv} (M)$ is not important. For this we can observe that 
   \bea
     \frac{1}{\Omega + d_v \alpha} = \frac{1}{\Omega} + d_v \left ( ... \right )
   \eea  
   and thus the equivariant class goes into another class. We need to check that the transformation \eqref{inverse-forms} is well-defined. $H_{\rm equiv} (M)$ is defined by the values of $\Omega_0$ at the fixed points and away from the fixed points the value of $\Omega_0$ can be shifted to any value by $d_v$-exact terms. If we deal with real valued $H_{\rm equiv} (M)$ and we choose $\Omega_0$ to have a different signs at different fixed points then we potentially have a problem since $\Omega_0$ will be zero somewhere between fixed points and its inverse is not well defined  \eqref{inverse-forms}. The way out is dictated by physics. We need to add a $\theta$-term to our observable which effectively complexifies $\Omega$ in the same way it would complexify the coupling of the gauge theory, $\frac{\theta}{2} + i \Omega$.  Now we can invert our new complexefied $\Omega$, moreover if we assume that $\Omega_0$ belongs to the upper half plane at a given fixed point (or to the lower half plane) then after S-duality $- (\Omega_0)^{-1}$ will belong again to the upper half plane at the same fixed point (or to the lower half plane correspondently). Thus under S-duality the observables which are parametrized by $H^{\mathbb C}_{\rm equiv} (M)$ (where we have to remove some purely imaginary lines) split chambers that are invariant under the action of S-duality.  To be more precise if we have an observable corresponding to a complexified $\Omega$ with $\Omega_0$ belonging to upper half plane at some fixed points and to lower half plane at  the remaining fixed points then after S-duality this distribution will not change.  Looking at the observable \eqref{cohomol-action} and   \eqref{cohomol-observ-ex} which corresponds to the supersymmetric ${\cal N}=2$ action for an abelian theory with the supersymmetry dictated by the choice of $s$ and $\tilde{s}$ and 
        \bea
         \Omega_0 = \frac{\tau s + \bar{\tau} \tilde{s}}{s+ \tilde{s}}~.
        \eea
     After cohomological S-duality we obtain
     \bea
      - \frac{1}{\Omega_0} = - \frac{s+ \tilde{s}}{\tau s + \bar{\tau} \tilde{s}} 
     \eea
      which is in the same cohomology class as 
      \bea
        \frac{-\frac{1}{\tau} s - \frac{1}{\bar{\tau}} \tilde{s}}{s+ \tilde{s}}
      \eea
       since 
       \bea
     \frac{s+ \tilde{s}}{\tau s + \bar{\tau} \tilde{s}} -  \frac{\frac{1}{\tau} s + \frac{1}{\bar{\tau}} \tilde{s}}{s+ \tilde{s}} = \frac{-(\tau - \bar{\tau})^2 s \tilde{s}}{\tau\bar{\tau} (\tau s + \bar{\tau} \tilde{s}) (s + \tilde{s})}
       \eea
   which vanishes at all fixed points $||v||^2 = s \tilde{s}$. Hence, cohomologically inverting $\Omega_0$ or inverting $\tau$ are the same and the treatment of S-duality from the previous subsection is consistent with the present cohomological discussion. 
  
  Let us make a brief comment about the contribution of the non-holomorphic fields to S-duality considerations.  In our logic we follow closely \cite{Losev:1997tp}. The non-holomorphic part of the vector multiplet enters through BRST-exact terms and is necessary to make the action positive definite. One can perform S-duality with additional BRST-exact terms (e.g, see the formulas for the non-equivariant case in  \cite{Losev:1997tp}) and the resulting formulas are not very inspiring. Upon certain field redefinitions the BRST-exact terms can be mapped to BRST-exact terms under S-duality. Since we have performed S-duality in the full supersymmetric theory in the previous subsection, there is no added value to repeat this fully in the cohomological variables. When we will discuss the non-linear case, we will come back to related issues.       
  
\section{{Non-linear ${\cal N}=2$ theory}}\label{section-4}\label{section-4}

 In this section we consider a non-linear ${\cal N}=2$ abelian supersymmetric theory in a supersymmetric rigid Sugra background and we discuss how S-duality acts (see also~\cite{Butter:2015tra} for a discussion of S-duality in this framework). We also briefly mention the cohomological description of gravitational corrections  to this theory. 
 
 \subsection{S-duality in the non-linear theory}\label{subsection-4.1}

Here we generalize the discussion in section~\ref{subsection-3.1} to apply to the non-linear case.  We consider several chiral multiplets $X^a$ of weight $w=1$. Given any holomorphic function ${\cal F}(X^a)$ which is homogenous of weight $w=2$ we can write down a supersymmetric Lagrangian as follows:
\begin{align}\label{lagdualc}
&-{i\over 2\pi}\left[\left( T^{({\cal F})}+{1\over 2} W^-_{\mu\nu}W^{-\mu\nu} {\cal F} \right) - \left( \b T^{({\cal F})}+{1\over 2} W^+_{\mu\nu}W^{+\mu\nu} \b {\cal F} \right)\right]~.
\end{align}
Here $T^{({\cal F})}$ is the top component of the chiral field which has ${\cal F}(X^a)$ as its lowest component (see~\eqref{holFmult}). For a holomorphic ${\cal F}(X^a)$ that is not homogenous of weight 2 we add an extra chiral multiplet $X^0$ of weight one (and an anti-chiral ${\bar X}^0$). We can then construct a holomorphic function ${\cal F}'(X^0,X^a)$ which is homogeneous of weight 2 and such that ${\cal F}'(1,X^a)={\cal F}(X^a)$. This ${\cal F}'$ can be used to write a supersymmetric Lagrangian as in \eqref{lagdualc}. Finally we can freeze the auxiliary chiral multiplet $X^0$ to the supersymmetric configuration~\eqref{bpschiral} (and similarly for ${\b X}^0$):
\bea\label{bpschtw}
&& X^0=1~,\qquad B^{+0}={\cal \bf F}^+- W^+~,\qquad D^0_{i j}= -2S_{i j}~,\cr
&& T^0= 2i(D^\mu+2i G^\mu) G_\mu+{1\over 2} W^-_{\mu\nu}\left(F^{-\mu\nu}-W^{-\mu\nu} \right)+\left({1\over 6}R-N\right)~.
\eea
In order to expand \eqref{lagdualc} in components it is useful to introduce $\widetilde {\cal F}= 2{\cal F} -{\cal F}_a X^a$ which vanishes for a homogenous ${\cal F}$ of weight $w=2$ and use  the following relations:
\bea\label{intrel}
&& \partial_{X^0} {\cal F}'\vert_{X^0=1}=2 {\cal F}-{\cal F}_a X^a=\widetilde {\cal F}~,\quad \partial_{X^0}\partial_{X^a} {\cal F}'\vert_{X^0=1}={\cal F}_a-{\cal F}_{a b} X^b=\widetilde {\cal F}_a~,\cr 
&& \partial^2_{X^0} {\cal F}'\vert_{X^0=1}= 2 {\cal F}-2{\cal F}_a X^a+{\cal F}_{a b} X^aX^b=\widetilde {\cal F}-\widetilde {\cal F}_a X^a~.
\eea

Next we introduce a vector multiplet for each of the chiral multiplets $X^a$ and write the coupling
\be\label{lagdualcc}
-{i\over 2\pi}\left[T( X^{a} X_{{\scriptscriptstyle D} a})+{1\over 2} W^-_{\mu\nu}W^{-\mu\nu} X^a X_{{\scriptscriptstyle D} a}-\b T (\b X^{a} \b X_{{\scriptscriptstyle D} a})-{1\over 2} W^+_{\mu\nu}W^{+\mu\nu} \b X^{a} \b X_{{\scriptscriptstyle D} a} \right]~.\nonumber
\ee
Note that we do not add vector multiplets that couple to the multiplets $X^0$ and $\b X^0$ that are frozen into a supersymmetric configuration.

Integrating over the vector multiplets enforces the constraints~\eqref{vectorconst}. This results in the following Lagrangian ($F^a=dA^a~,~~ g_{a b}=-i ({\cal F}_{a b}-\b {\cal F}_{ a b})$):

\be
\label{nonlinlag}
{\cal L}={1\over 4\pi} ({\cal L}_0+{\cal L}_1+{\cal L}_2)~.
\ee
Where the first term ${\cal L}_0$ is the minimal coupling to the metric and the $SU(2)_R$ background field of the flat space theory with prepotential ${\cal F}(X^a)$\footnote{Due to our choice of conventions (see appendix~\ref{sec:notations}) the sigma matrix $\sigma^{\mu\nu}$ is self-dual in the $\mu\nu$ indices. Hence ${\cal F}_{a b}$ multiplies $F_{\mu\nu}^{+a} F^{+b \mu\nu}$ which is not standard.},
\bea\label{prepveczero}
{\cal L}_{0}&=&{i\over 4} {\cal F}_{a b} F_{\mu\nu}^{+a} F^{+b \mu\nu}-{i\over 4} \b {\cal F}_{a b} F_{\mu\nu}^{-a} F^{-b \mu\nu}-g_{a b}\d^\mu{\b X^a}\, \d_\mu X^b +{1\over 8} g_{a b} D^{a i j} D^b_{i j}\cr&&-{1\over 2} {\cal F}_{a b}  \lambda^a_i \sigma^\mu D_\mu {\b \lambda}^{b i} +{1\over 2} \b {\cal F}_{a b} \bar \lambda^{a i}  \b \sigma^\mu D_\mu {\lambda}^b_i -{i\over 8} {\cal F}_{a b c} \lambda^{i a} \sigma^{\mu\nu}\lambda_i^{b} F^c_{\mu\nu}+{i\over 8} \b {\cal F}_{a b c} \b \lambda^{i a} \b \sigma^{\mu\nu}\b \lambda_i^{b} F^c_{\mu\nu}\cr&&+{i\over 8}{\cal F}_{a b c}  \lambda^{i a} \lambda^{j b}D^c_{i j}-{i\over 8}\b {\cal F}_{a b c}  \b\lambda^{i a} \b\lambda^{j b}D^c_{i j} -{i\over 48}{\cal F}_{a b c d}\lambda^{i a}\lambda^{j b}\lambda_i^c \lambda_j^d+{i\over 48}\b {\cal F}_{a b c d}\b\lambda^{i a}\b\lambda^{j b}\b\lambda_i^c \b \lambda_j^d~. \nonumber
\eea
The second term ${\cal L}_1$ includes couplings that are linear in supergravity auxiliary fields
\bea\label{prepvecone}{\cal L}_{1}&= &4G^\mu ({\cal F}_a \d_\mu \b X^a +\b {\cal F}_a \d_\mu X^a)-{i\over 2} ({\cal F}_a -\b {\cal F}_{a b} X^b)F^{a \mu\nu} W^-_{\mu\nu}+{i\over 2}\,({\b {\cal F}}_a-{\cal F}_{a b } \b X^b) F^{a \mu\nu} W^+_{\mu\nu}\cr
&&+{i\over 2} {\widetilde {\cal F}}_a ( D^a_{i j} S^{i j}+ F^{a \mu\nu} B^{+0}_{\mu\nu})-{i\over 2}\, {\widetilde {\b {\cal F}}}_a (D^a_{i j} S^{i j}+ F^{a\mu\nu} B^{-0}_{\mu\nu})\cr
&&+{i\over 4}{\cal F}_{a b c}  \lambda^{i a} \lambda^{j b}X^c S_{i j}-{i\over 4}\b {\cal F}_{a b c}  \b\lambda^{i a} \b\lambda^{j b}\b X^c S_{i j} +{1\over 4} g_{a b} G_\mu( \lambda^a_i  \sigma^\mu {\b \lambda}^{b i} -\bar \lambda^{a i}  \b \sigma^\mu {\lambda}^b_i)\cr
&&+{i\over 8} {\cal F}_{a b c} \lambda^{i a} \sigma^{\mu\nu}\lambda_i^{b} (X^c B^0_{\mu\nu}+ \b X^k W_{\mu\nu})-{i\over 8} \b {\cal F}_{a b c} \b \lambda^{i a} \b \sigma^{\mu\nu}\b \lambda_i^{b} (\b X^c B^0_{\mu\nu}+X^k W_{\mu\nu})~.~~~\qquad
\eea
Finally the last piece ${\cal L}_2$ is a potential for the scalars. It contains terms that are quadratic in supergravity auxiliary fields or that involve their derivatives. It also includes terms proportional to the combination  ${R\over 6}-N$ where $R$ is the Ricci scalar
\bea\label{prepvectwo}{\cal L}_{2}&= &- i(\widetilde {\cal F}+\!{\cal F}_a \b X^a\!-\widetilde{\b {\cal F}}-\!\b {\cal F}_a X^a) \left(\!{R\over 6}-\!N-4\,G^\mu G_\mu\! \right)+2 (\widetilde {\cal F}+\!{\cal F}_a \b X^a\!+\widetilde{\b {\cal F}}+\!\b {\cal F}_a X^a) \nabla^\mu G_\mu \cr 
&& +{i\over 4}(\widetilde {\cal F}-\widetilde {\cal F}_a X^a)(B^{+0}B^{+0}-2S^{i j}S_{i j})-{i\over 4}(2 {\cal F}-2 {\cal F}_a X^a+ \bar {\cal F}_{a b} X^a X^b)W^- W^-\cr
&&-{i\over 4}(\widetilde{\b {\cal F}}-\widetilde {\b {\cal F}}_a \b X^a)(B^{-0}B^{-0}-2 S^{i j}S_{i j})+{i\over 4}(2{\b {\cal F}}-2 \b {\cal F}_a \b X^a+{\cal F}_{a b} \b X^a\b X^b )W^+ W^+\cr
&&- {i\over 2} (\widetilde {\cal F} -\widetilde {\b {\cal F}_a} X^a) B^{-0} W^-+{i\over 2}(\widetilde{\b {\cal F}}-\widetilde {\cal F}_a \b X^a) B^{+0} W^+~.
\eea
The Lagrangian~\eqref{nonlinlag} is compatible with that appearing in~\cite{Butter:2015tra}. The differences stem from the use in~\cite{Butter:2015tra} of certain relations among the supergravity background fields that require the existence of eight separate supercharges.

In order to get the S-dual Lagrangian we proceed as in the quadratic case and integrate instead over  
$$T^a,\bar T^a, \psi^a,\bar \psi^a, B^{+a}, B^{- a} ,D^{a},\b D^{a}~.$$ 
This gives us the following
\bea\label{dualvrb}
&&{\cal F}_a= -X_{{\scriptscriptstyle D} a}~,\cr
&&\lambda_i^{a}= - {\cal F}^{a b} \lambda_{{\scriptscriptstyle D} b i}~,\cr
&& D^{a i j}= -{\cal F}^{a b}(D^{i j}_{{\scriptscriptstyle D} b} +\widetilde {\cal F}_b Y^{i j})+{1\over 2} {\cal F}_{a' b' c'}{\cal F}^{a' a}{\cal F}^{ b' b}{\cal F}^{c' c}\lambda_{{\scriptscriptstyle D} b}^i\lambda_{{\scriptscriptstyle D} c}^j~,\cr
&& B^{+a \mu\nu}=-{\cal F}^{a b}\left(dA_{{\scriptscriptstyle D} b}^{\mu\nu}-W^{+\mu\nu} \bar X_{{\scriptscriptstyle D}  b}+\widetilde {\cal F}_b B^{+0 \mu\nu}
\right)+{1\over 4}{\cal F}_{a' b' c'}{\cal F}^{a a'}{\cal F}^{b b'} {\cal F}^{c c'}\lambda_{{\scriptscriptstyle D} b}^{i}\sigma^{\mu\nu}\lambda_{{\scriptscriptstyle D}  c i}~.\qquad~~
\eea
By use of these relations, we obtain a Lagrangian for the vector multiplets that has the same form as~\eqref{nonlinlag} but with a prepotential $\widehat {\cal F}$ which is related to ${\cal F}$ by a Legendre transform.
\be
\widehat {\cal F}(X_{{\scriptscriptstyle D} a}) = {\cal F} + X_{ {\scriptscriptstyle D} a} X^a~,\qquad {\cal F}_a =-X_{{\scriptscriptstyle D} a}~.
\ee
The argument leading to $\hat {\cal F}$ is classical and quantum modifications are expected. These will be considered in section~\ref{subsection-5.2}

\subsection{Gravitational corrections}\label{subsection-4.2}

Since we consider a supersymmetric theory on a curved manifold one can construct  supersymmetric terms which involve derivatives of the background metric and other background supergravity fields. We refer to such terms as gravitational corrections. There are infinitely many such supersymmetric terms, e.g. the top components of
 \bea
  {\cal F}_g (X) W^{2g}~,
 \eea 
 where ${\cal F}_g$ is an arbitrary function and $W$ is the Weyl chiral superfield of ${\cal N}=2$ conformal supergravity ~\cite{Bergshoeff:1980is,Antoniadis:1993ze}. In the case of Donaldson-Witten theory there are two distinguished supersymmetric gravitational terms
  \bea
    \int f(\phi) \Tr (R \wedge \tilde{R})~,~~~~~\int g(\phi) \Tr(R \wedge R)~,\label{grav-DW}
  \eea 
 where up to normalization  $\Tr (R \wedge \tilde{R})$ corresponds to the Euler class and $\Tr(R \wedge R)$ to the signature class. If we switch to the equivariant Donaldson-Witten theory then (\ref{grav-DW}) are not supersymmetric since $\delta \phi \neq 0$. In the equivariant theory we are forced to choose an invariant metric and $\Tr (R \wedge \tilde{R})$ and $\Tr(R \wedge R)$ can be extended to equivariant characteristic classes: $\chi_{\rm equiv}$ and $\sigma_{\rm equiv}$.  Thus in the equivariant theory the terms  (\ref{grav-DW})  can replaced by the following 
    \bea
     \int f(\phi + \Psi + F) \chi_{\rm equiv}~,~~~~~~\int g(\phi + \Psi + F) \sigma_{\rm equiv}~. 
    \eea
These terms are examples of the observables \eqref{relation-diffs} with $\Omega$ being an equivariant characteristic class for the tangent bundle.  Hence it is natural to conjecture that up to BRST exact terms any gravitational correction can be written as function of the superfield ${\cal A}$ and the equivariant characteristic classes for the tangent bundle. Schematically we write
     \bea
      \int {\cal F} ({\cal A}, \chi_{\rm equiv}, \sigma_{\rm equiv})~. 
     \eea 
Depending on the geometry of $M$ we can switch to another basis of equivariant classes, e.g. to equivariant Chern classes for a complex manifold. 

\section{S-duality in cohomological variables}\label{section-5}\label{section-5}

 In this section we would like to study S-duality in cohomological variables in the context of a non-linear theory. First we  run some arguments from subsection \ref{subsection-3.2} and apply them to a non-linear abelian theory. Later we discuss the relation between S-duality and localization and we present some obstacle in treating S-duality as the Legendre transform. We argue that for S-duality to be compatible with localization we need to interpret S-duality as a Fourier transform. 
    
In this section the discussion is formal and it is applicable for any non-linear abelian supersymmetric theory. In the next section we discuss the implications for non-abelian supersymmetric Yang-Mills theory on a compact manifold. 
 
\subsection{Naive derivation}\label{subsection-5.1}

Let us consider a non linear  ${\cal N}=2$ theory for a collection of $U(1)$ vector multiplets on a manifold $M$. As we have explained the ${\cal N}=2$ vector mutiplets have a cohomological description in terms of the fields  $(A^a, \Psi^a, \phi^a, \varphi^a , \eta^a, \chi^a, H^a)$ where the label ``$a$'' runs over the collection of $U(1)$ multiplets. As before we concentrate on the holomorphic part of the multiplet which we combine in the superfelds  ${\cal A}^a = \phi^a + \Psi^a + F^a$. We can rewrite the nonlinear Lagrangian~\eqref{nonlinlag} using cohomological variables. This would result in the following observable (action) up to BRST-exact terms,
 \bea
     S=     \int      {\cal F}({\cal A}, \Omega)~,\label{general-nonlinear-obs}
   \eea
      which is invariant under the transformations 
    \bea
     \delta {\cal A} = (d + \iota_v) {\cal A}~.
    \eea
  provided that 
  \bea
     (d+ \iota_v) \Omega=(d+ \iota_v) \Big ( \Omega_0 + \Omega_2 + \Omega_4 \Big ) =0~.
  \eea    
  Here we may assume that ${\cal F}({\cal A}, \Omega)$ depends on a collection of equivariantly closed forms $\Omega$. However to avoid clutter we use just one form $\Omega$, the generalization to many $\Omega$'s being straightforward. The observable depends only on the equivaraint class of $\Omega$ since if  we change $\Omega$ by $d_v \alpha$ (provided that ${\cal L}_v \alpha =0$) we change the observable by a BRST exact term
     \bea
         \int      {\cal F}({\cal A}, \Omega+ d_v \alpha) =
          \int      {\cal F}({\cal A}, \Omega) + \delta \Big (    ... \Big )~. 
   \eea
   
  Now following the treatment from subsection \ref{subsection-3.2} we introduce two collections of  multiplets: long multiplets ${\cal A}^a$ and short multiplets ${\cal A}_\sD^a$ (see the formulas \eqref{long-multiplet} and \eqref{short-multiplet}). The action becomes 
      \bea
     S= \int  {\cal A}^a {\cal A}^b_\sD \delta_{ab}  +    {\cal F}({\cal A}, \Omega)
   \eea
    which in component is 
  \bea
  && S = \int D^a \phi_{\sD a} +  \rho^a \Psi_{\sD a} +  F^a F_{\sD a}  +   \Big [ \frac{\partial{\cal F}}{\partial \phi^a} D^a +
    \frac{\partial^2{\cal F}}{\partial \phi^a \partial \phi^b} (\Psi^a \rho^b + \frac{1}{2} F^a F^b ) \nn\\
   && +  \frac{1}{2}  \frac{\partial^3{\cal F}}{\partial \phi^a \partial \phi^b \partial \phi^c} \Psi^a \Psi^b F^c + \frac{1}{24} \frac{\partial^4{\cal F}}{\partial \phi^a \partial \phi^b \partial \phi^c \partial \phi^d} 
 \Psi^a \Psi^b \Psi^c \Psi^d \Big ] \nn \\
 && +   \Omega_2 \Big [\frac{\partial^2{\cal F}}{\partial \phi^a \partial \Omega_0} F^a + \frac{1}{2}    \frac{\partial^3{\cal F}}{\partial \phi^a \partial \phi^b \partial \Omega_0} \Psi^a \Psi^b \Big ] \nn\\
&& + \frac{\partial {\cal F}}{\partial \Omega_0} \Omega_4  + \frac{1}{2} \frac{\partial^2 {\cal F}}{\partial \Omega^2_0} \Omega_2 \Omega_2 ~,\label{BIG-dual-action}
  \eea
where ${\cal F} = {\cal F}(\phi, \Omega_0)$.  If we integrate out $\phi_\sD$, $\Psi_\sD$ and $F_\sD$  then the multiplets ${\cal A}^a$ shorten. In particular $\rho=0$, $D=0$ and $F$ becomes a curvature. Thus we obtain the observable \eqref{general-nonlinear-obs}
 \bea
&& \int {\cal F} (\phi + \Psi + F, \Omega_0 + \Omega_2 + \Omega_4) = \int  \frac{1}{2} \frac{\partial^2{\cal F}}{\partial \phi^a \partial \phi^b}  F^a F^b  \nn\\
   && + 
    \frac{1}{2}  \frac{\partial^3{\cal F}}{\partial \phi^a \partial \phi^b \partial \phi^c} \Psi^a \Psi^b F^c + \frac{1}{24} \frac{\partial^4{\cal F}}{\partial \phi^a \partial \phi^b  \partial \phi^c \partial \phi^d} 
 \Psi^a \Psi^b \Psi^c \Psi^d  \nn \\
 && +   \Omega_2 \Big [\frac{\partial^2{\cal F}}{\partial \phi^a \partial \Omega_0} F^a + \frac{1}{2}    \frac{\partial^3{\cal F}}{\partial \phi^a \partial \phi^b \partial \Omega_0} \Psi^a \Psi^b \Big ] \nn\\
&& + \frac{\partial {\cal F}}{\partial \Omega_0} \Omega_4  + \frac{1}{2} \frac{\partial^2 {\cal F}}{\partial \Omega^2_0} \Omega_2 \Omega_2 ~.
\eea
 Alternatively in \eqref{BIG-dual-action} we can integrate out $D$, $\rho$ and $F$ and obtain the following relations between fields
 \bea
 && \phi_{\sD a} +  \frac{\partial {\cal F}}{\partial \phi^a}=0~,\label{transform-1} \\
 && \Psi_{\sD a} +   \frac{\partial^2 {\cal F}}{\partial \phi^a \partial \phi^b} \Psi^b =0~,\\
 && F_{\sD a} + \frac{\partial^2  {\cal F}}{\partial \phi^a \partial \phi^b}F^b + \frac{\partial^3 {\cal F}}{\partial \phi^a \partial \phi^b \partial \phi^c} \Psi^b \Psi^c + \frac{\partial^2 {\cal F}}{\partial \phi^a \partial \Omega_0} \Omega_2 =0~. 
 \eea
By evaluating $S$ on this we get 
    \bea
    \int \hat{\cal F} (\phi_\sD + \Psi_\sD + F_\sD, \Omega_0 + \Omega_2 + \Omega_4) ~.
    \eea
    From \eqref{transform-1} we can  guess that we deal with the Legendre transform 
     \bea
      \phi^a   \phi_{\sD a} + {\cal F}(\phi, \Omega_0) = \hat{\cal F}(\phi_\sD, \Omega_0)~,\label{LT-1}
     \eea
    where we have assumed a $\Omega_0$-dependence  and thus we are dealing with a parametric Legendre transformation (see Appendix \ref{app:Legendre}). Let us introduce the following short-hand notations for the derivatives of ${\cal F}$
\bea
 \partial_{ab} {\cal F} = \frac{\partial^2 {\cal F}}{\partial \phi^a \partial \phi^b}~,~~~~
  \partial_{a0} {\cal F} = \frac{\partial^2 {\cal F}}{\partial \phi^a \partial \Omega_0}~,~~~~
   \partial_{0} {\cal F} = \frac{\partial {\cal F}}{\partial \Omega_0}~,~~~~
   \partial_{00} {\cal F} = \frac{\partial^2 {\cal F}}{\partial \Omega_0^2}~,
\eea
    and the following short-hand notations for the derivatives of the Legendre transform $\hat{\cal F}$
\bea
 \partial^{ab} \hat{\cal F} = \frac{\partial^2 \hat{\cal F}}{\partial \phi_{\sD a} \partial \phi_{\sD b}}~,~~~~
  \partial^{a0} \hat{\cal F} = \frac{\partial^2 \hat{\cal F}}{\partial \phi_{\sD a} \partial \Omega_0}~,~~~~
   \partial^{0} \hat{\cal F} = \frac{\partial \hat{\cal F}}{\partial \Omega_0}~,~~~~
   \partial^{00} \hat{\cal F} = \frac{\partial^2 \hat{\cal F}}{\partial \Omega_0^2}~.
\eea
 Following the logic presented in Appendix \ref{app:Legendre} we can derive  the following relations between different derivatives of 
  ${\cal F}$ and $\hat{\cal F}$
 \bea
     \partial^{ab} \hat{\cal F} (\phi_\sD, \Omega_0) =- \Big (  \partial_{ab} {\cal F} (\phi, \Omega_0) \Big)^{-1}|_{\phi = \phi(\phi_\sD, \Omega_0)}~, 
 \eea
 \bea
    \partial^{a0} \hat{\cal F} (\phi_\sD, \Omega_0) =    \partial^{ab} \hat{\cal F} (\phi_\sD, \Omega_0) 
  \Big (   \partial_{b0} {\cal F} (\phi, \Omega_0)\Big ) |_{\phi = \phi(\phi_\sD, \Omega_0)}~,
 \eea
\bea
 \partial^{00} \hat{\cal F} (\phi_\sD, \Omega_0) = \partial_{00} {\cal F}( \phi, \Omega_0) |_{\phi = \phi(\phi_\sD, \Omega_0)} + \partial^{a0} \hat{\cal F} (\phi_\sD, \Omega_0)  \partial_{a0} {\cal F}( \phi, \Omega_0) |_{\phi = \phi(\phi_\sD, \Omega_0)}~,
\eea
 where $\phi = \phi(\phi_\sD, \Omega_0)$ is obtained by inverting the formula \eqref{transform-1}. For the sake of clarity  let us concentrate only on the bosonic terms of \eqref{BIG-dual-action}
\bea
S= \int D^a \phi_{\sD a} + F^a F_{\sD a} + \partial_a {\cal F} D^a + \frac{1}{2} (\partial_{ab}{\cal F}) F^a F^b 
+ (\partial_{a0} {\cal F})  \Omega_2F^a + (\partial_0{\cal F} )\Omega_4 + \frac{1}{2} (\partial_{00} {\cal F}) \Omega_2^2~, \nn
\eea
 where we use our short-hand notations for the derivatives. Integrating out $D$ and $F$ we get the following relations
\bea
 &&\phi_{\sD a} + \partial_a {\cal F}=0~,\\
 && F_{\sD a} + \partial_{ab} {\cal F}~ F^b + \partial_{a0} {\cal F}~ \Omega_2 =0~,
\eea
which can be inverted
\bea
 F^a = \partial^{ab}\hat{\cal F} F_{\sD b} + \partial^{a0} \hat{\cal F} ~\Omega_2~.
\eea
Evaluating $S$ on this we will get 
\bea
\label{fingt}
\int \frac{1}{2} (\partial^{ab}\hat{\cal F}) F_{\sD a} F_{\sD b} 
+ (\partial^{a0} \hat{\cal F})  \Omega_2F_{\sD a} + (\partial^0 \hat{\cal F} )\Omega_4 + \frac{1}{2} (\partial^{00} \hat{\cal F}) \Omega_2^2~,
\eea
 where we used formulas for parametric Legendre transforms reviewed in appendix~\ref{app:Legendre}.  The fermionic terms work similarly but the manipulations required are more involved.  We have generalized the treatment of S-duality  to a cohomological non-linear observable. The present cohomological discussion is compatible with the derivation presented in subsection \ref{subsection-4.1}. The argument leading to~\eqref{fingt} is classical. In the next section we will see how it gets modified.

\subsection{S-duality as Fourier transform}\label{subsection-5.2}

We have seen that the equivariant version of S-duality appears to relate the observable corresponding to ${\cal F}$ to that corresponding to its Legendre transform $\hat {\cal F}$. However this is incompatible with localization.
Consider a simply connected manifold equipped with the data that we have described in subsection \ref{subsection-2.1} (Killing vector field $v$ with isolated fixed point etc.). We are interested in calculating the partition function on $M$ for a non-linear $U(1)$ ${\cal N}=2$ gauge theory with action
\bea
  \int {\cal F} ({\cal A}, \Omega)~.\label{formal-observable} 
\eea
 The application of localization to this theory proceeds along the lines discussed in \cite{Festuccia:2018rew}. We have to add BRST-exact terms involving non-holomprhic fields. Since this theory is abelian, we expect that the answer is simpler compared with that for a non-abelian theory. The path integral will get two types of contributions, point like instantons and fluxes controlled by $H^2(M, \mathbb{Z})$.  For the sake of clarity let us assume that we have just one $U(1)$ vector multiplet. Since we deal with a $U(1)$ theory the contributions of the point-like instantons depend only on  the local toric data at every fixed point and  are universal for any ${\cal F}$.  Ignoring these universal contributions the partition function for the theory has the following structure
      \bea
      Z \sim \sum\limits_{k_i} \int da ~~e^{2\pi \sum\limits_{i} \frac{1}{\epsilon_i \epsilon_i'}{{\cal F}\Big (ia + k_i, \Omega_0(x_i)\Big ) }}~,\label{PF-F-abelian}
     \eea  
where $\phi(x_i) = i a + k_i$ with $x_i$ being fixed points and $k_i$ being discrete data that corresponds to the fluxes (these are discrete shifts which involve also  fixed point data, later we give an explicit example for $\mathbb{CP}^2$).  Here we also ignored possible one-loop contributions that can be brought out of the integral since the theory is abelian. These contributions are also universal, i.e. they are independent from the form of ${\cal F}$.  
      
  In the previous subsection we have shown that classically the theory corresponding to ${\cal F}$ should be equivalent to an S-dual theory with $\hat{\cal F}$ that is the parametric Legendre transform of ${\cal F}$. If we also localized this S-dual theory we would obtain the following partition function 
       \bea
      Z \sim \sum\limits_{k_i} \int da ~~e^{2\pi \sum\limits_{i} \frac{1}{\epsilon_i \epsilon_i'}{\hat{\cal F}\Big (ia + k_i, \Omega_0(x_i)\Big ) }}~.\label{PF-F-abelian-dual}
     \eea 
 The expressions~\eqref{PF-F-abelian} and \eqref{PF-F-abelian-dual} should coincide if corresponding to the same theory.  Moreover this should be true for any choice of function ${\cal F}$ and its Legendre transform $\hat{\cal F}$. However the two integrals are not the same, hence the classical result needs to be modified. Indeed we will show that it is consistent for  $e^{\cal F}$ and $e^{\hat{\cal F}}$ to be related through a Fourier transform. In the limit of small $\epsilon$'s the Fourier transform gives rise to the Legendre transform. 
         
         The following is a heuristic argument for why S-duality in the equivariant setting should correspond to a Fourier transform.  Consider an action for one long and one short multiplet       
 \bea
     S= \int  {\cal A} {\cal A}_\sD  +    {\cal F}({\cal A}, \Omega)~.\label{FT-local-action}
   \eea
This action is written  in components in \eqref{BIG-dual-action} (here for the clarity we deal just with one $U(1)$). Now we can try to apply  localization in the presence of  both multiplets by adding appropriate BRST exact terms. The action \eqref{FT-local-action} is invariant under the following supersymmetry 
  \bea
   \delta {\cal A} = (d + \iota_v) {\cal A} ~, ~~~~  \delta {\cal A}_\sD = (d + \iota_v) {\cal A}_\sD~.
  \eea
Using the standard localization logic we can add to the action \eqref{FT-local-action} the following BRST-exact terms using an invariant metric
 \bea
   \int \delta ({\cal A} \wedge \star \overline{\delta{\cal A}} + {\cal A}_\sD \wedge \star \overline{\delta{\cal A}_\sD})  ~,
 \eea
  which in component looks as
      \bea
       || \iota_v F + d\phi ||^2 +  || \iota_v D + dF||^2 + ||\iota_v F_\sD + d\phi_\sD || + ... ~,\label{BRST-exact-bos-naive}
        \eea
         where dots stand for the fermionic terms.  Thus on the localization locus we can evaluate the action  \eqref{FT-local-action}  using equivariant localization
   \bea      
 S= 2\pi \sum\limits_i \frac{1}{\epsilon_i \epsilon_i'} \Big ( \phi(x_i) \phi_\sD(x_i) +  {\cal F} (\phi(x_i), \Omega_0(x_i) \Big ) ~,
\eea
 where we sum over all fixed points $x_i$.  In the path integral we would expect that we integrate over allowed values of $\phi(x_i)$  and $\phi_\sD(x_i)$.  Thus in the localized variables we get a Fourier transform instead of a Legendre transform. Also using the ideas from subsection \ref{subsection-2.3} we can formally derive above result. 
  
The discussion above is heuristic. The BRST exact terms \eqref{BRST-exact-bos-naive} admit huge kernels and this should be fixed. Both for the short multiplet (which is part of a ${\cal N}=2$ vector multiplet)  and for the long multiplet (which is part of a ${\cal N}=2$ chiral multiplet) we should add all remaining fields and construct  positive BRST-exact terms with at most finite dimensional kernels. The goal of these additional BRST-exact terms is  to pick up a reasonable representative. The analysis of these additional terms leads to rather messy PDEs which we  find hard to analyse. For the vector multiplet (short multiplet ${\cal A}_\sD$) the relevant analysis was presented in \cite{Festuccia:2018rew}.  There we argued that $\phi_\sD(x_i) = i a + k_i$, where $a$ is constant and $k_i$'s correspond  to discrete flux contributions. For the vector multiplet there will be point like instantons and one-loop contributions which are    universal and independent of ${\cal F}$. One should perform a similar analysis for the chiral multiplet (long multiplet plus additional fields).  At the moment we are unable to perform a consistent analysis of BRST-exact terms and corresponding PDEs for all the fields in  the chiral multiplet. However we expect that our previous heuristic analysis gives the right result. Thus schematically localizing with both multiplets present we get the following expression
\bea
 Z \sim \sum\limits_{k_i}  \int \prod\limits_i d\phi_i \int da~ e^{2\pi \sum\limits_{i} \frac{1}{\epsilon_i \epsilon_i'} \Big ( \phi_i (ia + k_i) +
 {{\cal F} (\phi_i,  \Omega_0(x_i) ) \Big ) }}~,\label{FT-conjecture}
\eea  
 where we use $\phi_i = \phi(x_i)$. Here we ignore the contributions of point like instantons for the vector multiplets,  and one loop factors for  vector and chiral multiplets which can be brought outside of the integral. The  suggested formula \eqref{FT-conjecture} is conjectural. Moreover there are ambiguities in choosing the integrating contour  over $\phi_i$ and possibly over $a$. 
  As well there can be ambiguities related to the normalization of the Fourier transform which may come from the proper treatment of the zero modes. 
 In the next subsection we will consider two explicit examples of this formula that hopefully can bring some  clarification. 

 Let us finish this subsection with some  remarks about the cohomological difference between Legendre transform and  Fourier transform. If we look at the answer \eqref{PF-F-abelian} and ignore fluxes we see that it is given in terms of  a function ${\cal F}(a, \Omega_0(x_i))$ that depends on different parameters $\Omega_0(x_i)$ at different fixed points.  Since $\Omega_0(x)$ is the zero form component of an equivariantly closed form it makes sense to consider an object  ${\cal F}(a, \Omega)$ where $\Omega$ is an equivariantly closed form.  We understand ${\cal F}(a, \Omega)$ as a differential form 
       \bea
  {\cal F}(a, \Omega) = {\cal F}(a, \Omega_0) + (\Omega_2 + \Omega_4) \frac{\partial}{\partial \Omega_0} {\cal F} (a, \Omega_0) + 
   \frac{1}{2} \Omega_2^2  \frac{\partial^2}{\partial \Omega_0^2} {\cal F} (a, \Omega_0)~,
 \eea
  which is not uniquely defined since we are interested only in its class. Thus we have the following identification 
   \bea
    {\cal F}(a, \Omega) - {\cal F}(a, \Omega') = d_v (...)~,\label{formal-equivalence} 
   \eea
   that involves an equivariant differential $d_v$ (here we also assume that ${\cal L}_v(...)=0$ since we use the Cartan model of equivarant cohomology).  This property guarantees that ${\cal F}(a, \Omega_0(x_i))$ remains unchanged.   At the level of the cohomological observable it implies that 
 \bea
  \int {\cal F} ({\cal A}, \Omega) -  \int {\cal F} ({\cal A}, \Omega') = \delta (...)~. 
\eea

Consider now the Legendre transform $\tilde{\cal F}$ of a function ${\cal F}$. We can again write the expansion
  \bea
  \tilde{\cal F}(a, \Omega) = \tilde{\cal F}(a, \Omega_0) + (\Omega_2 + \Omega_4) \frac{\partial}{\partial \Omega_0} \tilde{\cal F} (a, \Omega_0) + 
   \frac{1}{2} \Omega_2^2  \frac{\partial^2}{\partial \Omega_0^2} \tilde{\cal F} (a, \Omega_0)~. 
 \eea
  Since the second derivative of $\tilde{\cal F}$ with respect to a parameter $\Omega_0$ transforms in a complicated way then from \eqref{formal-equivalence}
   we get
  \bea
  \tilde{\cal F}(a, \Omega) - \tilde{\cal F}(a, \Omega') \neq d_v (...)~. 
 \eea
  Thus the Legendre transform does not respect the cohomological identification. The  Fourier transform behaves quite differently.  The relation \eqref{formal-equivalence}
   implies 
   \bea
    e^{{\cal F}(a, \Omega)} - e^{{\cal F}(a, \Omega')} = d_v (...)~.\label{exp-formal-equivalence} 
   \eea
 Define the formal Fourier transform\footnote{From now on we use $e^{\hat{\cal F}}$ to denote the Fourier transform.}  as follows
\bea
 &&\int d a ~  e^{iaa_\sD + {\cal F}(a_\sD, \Omega)} \nn \\
 && =   \int d a ~  e^{iaa_\sD} \Big ( e^{{\cal F}(a, \Omega_0)} + \frac{\partial}{\partial \Omega_0} e^{{\cal F}(a, \Omega_0)}  \Omega_2 +  \frac{\partial}{\partial \Omega_0} e^{{\cal F}(a, \Omega_0) } \Omega_4 + \frac{1}{2}  \frac{\partial^2}{\partial \Omega_0^2} e^{{\cal F}(a, \Omega_0) } \Omega_2^2 \Big ) \nn\\
&& =  \Big (1 + (\Omega_2 + \Omega_4) \frac{\partial}{\partial \Omega_0} + \frac{1}{2} \Omega_2^2 \frac{\partial^2}{\partial \Omega_0^2} \Big )
 \int d a ~  e^{iaa_\sD + {\cal F}(a, \Omega_0)} = e^{\hat{\cal F} (a, \Omega)}
 \eea
 Then the relation \eqref{exp-formal-equivalence} would imply that  
 \bea
  e^{\hat{\cal F}(a, \Omega)} - e^{\hat{\cal F}(a, \Omega')} = d_v (...)~. 
 \eea
  Let us stress that this observation does not constitute a proof and does not contradict the field theoretical considerations presented in the previous subsection.  There we guaranteed supersymmetry working ``on-shell''. Here we want to point out   that if we deal with functions that depend on equivariant forms and some variable $a$ then the Fourier transform is better suited to work with equivariant cohomology classes. Although the Legendre transform is the leading semi-classical approximation in $\hbar$  to the Fourier transform  we do not see any canonical way to deal with $\hbar$ for $d_v$-exact terms. We will comment more on the relation between the Legendre transform  and the Fourier transform in the present context in section~\ref{section-6}. 
 
 \subsection{Examples}   
 
 Let us consider two examples which will clarify formula  \eqref{FT-conjecture} and the interpretation of S-duality as Fourier transform. In these examples we deal with various analytical issues (e.g., choice of integration contour, delta functions etc) in a rather formal fashion. It may happen that S-duality may fix some of these issues,   for example the contour of integration should be chosen such that various manipulations actually work. Another important comment is that the presence of the fluxes   plays a crucial role in the reducing the multiple integrals in  \eqref{FT-conjecture} to either \eqref{PF-F-abelian} or \eqref{PF-F-abelian-dual}. One can actually perform a simple  count of variables and conclude that for these manipulations to work in principle we need that  the number of fixed points minus two  should be equal to the  number of fluxes which is exactly $H^2(M, \mathbb{Z})$ on simply connected manifolds with $T^2$-action. We hope to clarify this point with the example of ${\mathbb CP}^2$. 
 
 \subsubsection{$S^4$}
 
 Let us start from the simple example of $S^4$ which does not involve fluxes since $H^2$ is empty. If we take a non-linear $U(1)$ theory with the observable given by   ${\cal F}({\cal A}, \Omega)$ then up to overall universal factors the answer is given by the following integral
  \bea
   Z_{S^4} = \int da~ e^{\frac{2\pi}{\epsilon_1 \epsilon_2} \Big ({\cal F}(ia, \Omega_0(x_N)) - {\cal F}(ia, \Omega(x_S) ) \Big )}~,\label{S4-integral-abelian} 
  \eea
   where $\epsilon_1$, $\epsilon_2$ are equivariant parameters and $x_N, x_S$ are the two fixed points. The minus sign follows from the identification of the equivariant parameters and the choice of contour can be motivated by the reality conditions coming from the action, (see \cite{Festuccia:2018rew} for further explanations).  Using the conventions for the Fourier transform from Appendix \ref{app:Fourier} we can rewrite (\ref{S4-integral-abelian}) as follows
      \bea
   Z_{S^4} = \frac{1}{| \epsilon_1 \epsilon_2 |}  \int da ~d\phi_1~ d\phi_2 ~ e^{\frac{2\pi}{\epsilon_1 \epsilon_2} \Big (ia (\phi_1 - \phi_2) + \hat{\cal F}(i \phi_1, \Omega_0(x_N)) - \hat{\cal F}(i\phi_2, \Omega(x_S) ) \Big )}~.
  \eea
   Since the original integral over $a$ is performed along the imaginary line $i\mathbb{R}$ then it is natural to chose  the same contours for the $\phi_1$ and $\phi_2$ integrals. The integration over $a$ brings us back to the expression \eqref{S4-integral-abelian} but with the Fourier transformed $\hat{\cal F}$.  Thus S-duality works for $S^4$ in rather straightforward way. 
  
Referring to the setup of the original  work of Pestun on $S^4$ \cite{Pestun:2007rz}  (and extended for the squashed $S^4$ in \cite{Hama:2012bg, Pestun:2014mja}) we consider a non-linear $U(1)$ theory  and choose the observable built from $2\pi  {\cal F}_{\rm Nekr}= \epsilon_1 \epsilon_2 \log Z_{\rm Nekr} $ related to the Nekrasov partition function $Z_{\rm Nekr}$ on $\mathbb{C}^2$.  We have to choose $\Omega$ appropriately (see next section for further discussion). Then for this non-linear $U(1)$ theory  the partition function is given by  
    \bea
    Z_{S^4} = ||~e^{\frac{2\pi}{\epsilon_1\epsilon_2} {\cal F}_{\rm Nekr}} ~||^2 ~, 
   \eea
  where $||f(a)||$ denotes the $L^2$ norm. Performing the Fourier transform on $a$ the answer does not change since 
   \bea
    ||~e^{\frac{2\pi}{\epsilon_1\epsilon_2} {\cal F}_{\rm Nekr}} ~||^2 =  ||~e^{\frac{2\pi}{\epsilon_1\epsilon_2} \hat{\cal F}_{\rm Nekr}} ~||^2~,
   \eea
 which is the well-known Plancherel theorem for the Fourier transform.  
 
 As we have argued in \cite{Festuccia:2018rew} we expect that the answer for $Z_{S^4} $ should be holomorphic in $\epsilon$'s. The manipulations with the Fourier transform above, however require real $\epsilon$'s and introduce normalization factors involving absolute values. We do not analyze analytic properties in $\epsilon$'s here. The final result could be extended analytically away from real $\epsilon$'s.  The same comments are applicable to the next example. 
 
 \subsubsection{$\mathbb{CP}^2$} 

The next example we consider is the complex projective space $\mathbb{CP}^2$ with the standard $T^2$-action. With respect to this action $\mathbb{CP}^2$ has three fixed points $x_i$ ($i=1,2,3$). The equivariant parameters corresponding to each fixed point are related as
 \bea
 (\epsilon_1, \epsilon_2)~,~~~~~
 (\epsilon_2-\epsilon_1, -\epsilon_1)~,~~~~~
 (\epsilon_1 - \epsilon_2, - \epsilon_2)~,
 \eea
 which follows from considering the standard homogeneous coordinates. Consider the supersymmetric non-linear $U(1)$ gauge theory determined by ${\cal F}$. As  explained in \cite{Festuccia:2018rew} $\mathbb{CP}^2$ admits different supersymmetries  related to different assignments of $\pm$ labels for each fixed point but we treat all cases uniformly.  Up to universal factors that multiply the overall answer the partition function can be written as follows
 \bea
Z_{\mathbb{CP}^2} = \label{formula-1} \sum\limits_{n \in \mathbb{Z}} \int d a~ e^{\frac{2\pi {\cal F}_1(i a + p \epsilon_1 + q \epsilon_2 ) }{\epsilon_1 \epsilon_2} + \frac{2\pi  {\cal F}_2(i a + q(\epsilon_2 - \epsilon_1) + r (-\epsilon_1)) }{(\epsilon_2 - \epsilon_1) (-\epsilon_1)}  + \frac{2\pi  {\cal F}_3(i a + p(\epsilon_1 - \epsilon_2) + r (-\epsilon_2))}{(\epsilon_1 - \epsilon_2) (-\epsilon_2)} }  
\eea
 where we use the following short hand notation ${\cal F}_i (a) = {\cal F} (a, \Omega_0(x_i))$.  Here for each $n\in \mathbb{Z}$ the integers $(p,q,r)$   are such that $p+q+r=n$. They are introduced so that the three fixed points appear in expression \eqref{formula-1} on an equal footing.  We will see below that the final answer should not depend on the specific choice of $(p,q,r)$.
 
 Following  \cite{Festuccia:2018rew} we should analyze the equivariant condition
 \bea
  \iota_v F + d\phi =0~.
 \eea
Imposing that  the integral over a two cycle of $F$ is quantized it follows that the real part of $\phi(x_i)-\phi(x_j)$ is quantized (here $x_i$, $x_j$ are fixed points and $i\neq j$).  Formula \eqref{formula-1} is symmetric in shifts of $a$, however if we  perform a formal shift of the contour
    \bea
     i\tilde{a} = i a + p \epsilon_1 + q \epsilon_2 
    \eea
    then formula \eqref{formula-1} becomes 
    \bea
 Z_{\mathbb{CP}^2} = 
  \sum\limits_{n \in \mathbb{Z}} \int d \tilde{a}~ e^{\frac{2\pi}{\epsilon_1 \epsilon_2} {\cal F}_1 (i \tilde{a} ) + \frac{2\pi }{(\epsilon_2 - \epsilon_1) (-\epsilon_1)}  
 {\cal F}_2 (i \tilde{a} - \epsilon_1 n )
 + \frac{2\pi }{(\epsilon_1 - \epsilon_2) (-\epsilon_2)}  
 {\cal F}_3 (i \tilde{a} -\epsilon_2 n)} ~. \label{formula-CP-new}
\eea
 We would like to stress that we do not understand how to choose the contour of integration from first principles. Following the considerations from \cite{Pestun:2007rz} and \cite{Festuccia:2018rew} we may deduce the appropriate contour from the the reality conditions on the physical fields. 
 
Putting aside the problem of  choosing the  contour, let us proceed formally and introduce the following Fourier transforms 
 \bea
 e^{\frac{2\pi}{\epsilon_1 \epsilon_2} {\cal F}_1 (i a + p \epsilon_1 + q \epsilon_2 )} = \frac{1}{\sqrt{|\epsilon_1 \epsilon_2|}}     \int d\phi_1~ e^{\frac{2\pi}{\epsilon_1\epsilon_2} \Big [ \phi_1 (i a + p \epsilon_1 + q \epsilon_2 ) +
 \hat{\cal F}_1 (i \phi_1) \Big ]}~,\nn
  \eea
\bea
e^{ \frac{2\pi }{(\epsilon_2 - \epsilon_1) (-\epsilon_1)}  
 {\cal F}_2 (i a + q(\epsilon_2 - \epsilon_1) + r (-\epsilon_1))} = \frac{1}{\sqrt{|(\epsilon_2 - \epsilon_1)\epsilon_1|}}
   \int d\phi_2~ e^{ \frac{2\pi}{(\epsilon_2 - \epsilon_1) (-\epsilon_1)} \Big [ \phi_2 (ia  +
 q(\epsilon_2 - \epsilon_1) + r (-\epsilon_1)) + \hat{\cal F}_2(i\phi_2) \Big ] }~,\nn
\eea
\bea
 e^{\frac{2\pi }{(\epsilon_1 - \epsilon_2) (-\epsilon_2)}  
 {\cal F}_3 (i a + p(\epsilon_1 - \epsilon_2) + r (-\epsilon_2))} = \frac{1}{\sqrt{|(\epsilon_2 - \epsilon_1) \epsilon_2|}}
  \int d\phi_3 ~ e^{\frac{2\pi }{(\epsilon_1 - \epsilon_2) (-\epsilon_2)} \Big [ \phi_3 
  (i a + p(\epsilon_1 - \epsilon_2) + r (-\epsilon_2)) + \hat{\cal F}_3 (i \phi_3) \Big ] }~,\nn
\eea
 which we substitute into formula \eqref{formula-1}. Integrating over $a$ we get the following delta function
 \bea
   \delta \Big ( \frac{\phi_1}{\epsilon_1\epsilon_2} +\frac{\phi_2}{(\epsilon_2 - \epsilon_1) (-\epsilon_1)} + \frac{\phi_3}{(\epsilon_1 - \epsilon_2) (-\epsilon_2)} \Big )~,
 \eea
 which removes the integration over $\phi_3$ (and cancels the factor $|(\epsilon_2 - \epsilon_1) \epsilon_2|$) and collecting terms with discrete shifts we isolate a factor of
\bea
\frac{1}{|\epsilon_1|}  \sum\limits_{n \in \mathbb{Z}}  e^{\frac{2\pi n}{\epsilon_1} (\phi_1 - \phi_2)}~.
\eea
  We interpret this as a periodic delta function which imposes the the constraint
 \bea
  \frac{1}{\epsilon_1} (\phi_1 - \phi_2) \in i \mathbb{Z}~.
 \eea
 We can solve the constraint as
  \bea
   \phi_1 = \lambda~,~~~~~
   \phi_2 = \lambda + i \epsilon_1 k~,~~~~~
   \phi_3 = \lambda + i \epsilon_2 k
  \eea
   with $k \in \mathbb{Z}$ and $\lambda \in {\mathbb R}$.  Finally substituting back we get 
   \bea
    Z_{\mathbb{CP}^2} = \sum\limits_{k \in \mathbb{Z}} \int d \lambda~ e^{\frac{2\pi}{\epsilon_1\epsilon_2} \hat{\cal F}_1 (i \lambda) 
     +   \frac{2\pi}{(\epsilon_2 - \epsilon_1)(-\epsilon_1)}          \hat{\cal F}_2 (i\lambda - \epsilon_1 k) 
      + \frac{2\pi }{(\epsilon_1 - \epsilon_2) (-\epsilon_2)}  \hat{\cal F}_3 (i \lambda - \epsilon_2 k)}~,
   \eea
    which is in agreement with expression \eqref{formula-CP-new}. As we can see the fluxes play a very crucial role in the argument. Similar formal manipulations will work in examples with more fixed points. For simply connected $M$ the number of fluxes is the dimension of $H^2$ which is number of fixed points minus two. This will guarantee that an argument parallel to that for $\mathbb{CP}^2$ will work.   We believe that the analytical issues should be resolved in such fashion that S-duality is implemented as described.  In this example, as for  $S^4$, we restrict the  $\epsilon$'s to be real. As for the example of $S^4$ we can replace ${\cal F}$ with copies of the Nekrasov function $    \frac{\epsilon_1 \epsilon_2}{2\pi} \log Z_{\rm Nekr}$ with the appropriate identifications. In the next section we will comment on the meaning of these manipulations.

\section{Effective ${\cal N}=2$ abelian theory}\label{section-6}

Let us get back to considering a non-abelian ${\cal N}=2$ supersymmetry gauge theory on compact manifold. As stressed earlier, we do not know how to derive the partition function from first principles. However we conjecture that the answer should have the following form 
  \bea
      Z = \sum\limits_{k_i} \int da ~~e^{2\pi \sum\limits_{i =1}^{p} \frac{1}{\epsilon_i \epsilon_i'}{\cal F}^{\rm ins}_{\rm Nekr} \Big (ia + k_i, \Lambda, \epsilon_i, \epsilon_i' \Big )
      + 2\pi \sum\limits_{i=p+1}^l \frac{1}{\epsilon_i \epsilon_i'}{\cal F}^{\rm anti-ins}_{\rm Nekr} \Big (ia + k_i, \bar{\Lambda}, \epsilon_i, \epsilon_i' \Big ) }~,\label{sect6-PF-conj-intro}
     \eea 
 where we count point like instantons at some fixed points, point like anti-instantons at other fixed points and fluxes controlled by $H^2$. In light of our previous discussion of S-duality in the non-linear $U(1)$ theory the following natural question arises: can we construct some non-linear $U(1)$ theory which gives exactly the same localization answer as \eqref{sect6-PF-conj-intro}?  Below we present a construction of this theory by reverse engineering from the conjectured answer. As we don't have an alternative derivation one should be critical of our considerations. In order to be able to get~\eqref{sect6-PF-conj-intro} from a non-linear $U(1)$ theory we should promote the parameters of the theory to equivariant classes. This is somewhat similar to the idea of promoting the parameters of many supersymmetric theories  to the expectation values of some background superfields. 
 
Before suggesting the answer we have to do some preparatory work. Let us introduce some equivariant classes. Generalizing the formulas \eqref{cohomol-observ-ex} we can define the following family of equivariant closed forms $\Omega (a, b) = \Omega_0 (a,b) + \Omega_2 (a, b)  + \Omega_4 (a,b)$ which depend on  two complex parameters $a,b \in {\mathbb C}$ 
  \bea
  && \Omega_0 (a, b) = \frac{a s + b \tilde{s}}{s+\tilde{s}}~,\nn\\
  && \Omega_2 (a, b) =  (b-a)  \frac{s-\tilde{s}}{(s+ \tilde{s})^3} dk  + 2 \frac{b-a}{(s+ \tilde{s})^3} \kappa \wedge d (s-\tilde{s})~,\label{def-omega-general}\\
  && \Omega_4 (a, b) = 3 (a-b) \frac{s-\tilde{s}}{(s+\tilde{s})^5} d\kappa \wedge d \kappa +  12 \frac{a-b}{(s+\tilde{s})^5} \kappa \wedge d\kappa \wedge d (s - \tilde{s})~.\nn
  \eea      
 One ca check that these satisfy $d_v \Omega (a,b)=0$  where $d_v = d+ \iota_v$ using  $\iota_v \kappa= ||v||^2 = s \tilde{s}$ and ${\cal L}_v \kappa =0$.  We can exponentiate this class and define the following equivariant class 
 \bea
  e^{2\pi i \Omega} = e^{2\pi i \Omega_0} ( 1 + 2\pi i \Omega_2 + 2\pi i \Omega_4 - 2\pi^2 \Omega_2^2)~. 
 \eea        
 Following standard notation we can introduce 
 \bea
  \Lambda=  e^{2\pi i \tau}~,~~~~~\bar{\Lambda} =  e^{-2\pi i \bar{\tau}}~,
 \eea
  where we ignore the mass scale. $\Lambda$ ($\bar{\Lambda}$) are the standard instanton (anti-instanton) counting parameters. We can define an equivariant class as 
 \bea
  {\mathbf \Lambda} = e^{2\pi i \Omega(\tau, - \bar{\tau})} = \Lambda^{\frac{s}{s+ \tilde{s}}} \bar{\Lambda}^{\frac{\tilde{s}}{s+\tilde{s}}} (1 + ... ) ~, \label{defin-Lambda-class}
 \eea
  its lowest component at fixed points will be either $\Lambda$ (when $s=1$ and $\tilde{s}=0$) or $\bar{\Lambda}$ (when $\tilde{s}=1$ and 
   $s=0$).  
 
 Let us consider the Nekrasov partition function for  a pure vector multiplet on ${\mathbb C}^2$. We use the following conventions
  \bea
  Z^{\rm inst}_{\rm Nekr} (a, \epsilon_1, \epsilon_2, \Lambda) = e^{\frac{2\pi}{\epsilon_1 \epsilon_2} {\cal F}^{\rm inst}_{\rm Nekr} (a, \epsilon_1, \epsilon_2, \Lambda)}~,
 \eea 
   where $Z^{\rm inst}_{\rm Nekr}$ is assumed to contain everything including classical  and 1-loop contributions.   We assume the following symmetry
  \bea
    Z^{\rm inst}_{\rm Nekr} (a, -\epsilon_1, - \epsilon_2, \Lambda) =  Z^{\rm inst}_{\rm Nekr} (a, \epsilon_1, \epsilon_2, \Lambda)~.
    \label{symmetry-flat}
  \eea
 This corresponds to $z_1 \rightarrow \bar{z}_1$, $z_2 \rightarrow \bar{z}_2$ to which the vector multiplet should be blind.  On $\mathbb{C}^2$  this is true for the classical contribution and for the instanton contribution but the perturbative part is not invariant under $(\epsilon_1, \epsilon_2)\rightarrow (-\epsilon_1, -\epsilon_2)$.  On a compact manifold, this symmetry is restored upon gluing the perturbative parts corresponding to the various fixed points \footnote{We think that the anomalous property of the perturbative contribution on $ \mathbb{C}^2$ under the symmetry  $(\epsilon_1, \epsilon_2)\rightarrow (-\epsilon_1, -\epsilon_2) $ has a cohomological origin and should be better studied in the appropriate language.}. In general we have the following expansion for the Nekrasov function
 \bea
  \log Z^{\rm inst}_{\rm Nekr} (a, \epsilon_1, \epsilon_2, \Lambda) = \sum_{i,j=0}^\infty (\epsilon_1 + \epsilon_2)^i (\epsilon_1 \epsilon_2)^{j-1} F^{(\frac{i}{2}, j)} ~.
 \eea
  Assuming that the property \eqref{symmetry-flat} holds for the full Nekrasov function we can rearrange the sum above in terms of the equivariant Euler characteristic and of the equivaraint signature
   \bea
   \chi_{\rm eq} = \epsilon_1 \epsilon_2~,~~~~~\sigma_{\rm eq} = \frac{\epsilon_1^2 + \epsilon_2^2}{3}~, 
  \eea
   which can be promoted to equivariant classes.  Thus we can think of the Nekrasov function as a function of characteristic classes. In a way this is not surprising since direct calculations of the Nekrasov function are related to index theorem calculations. Even the perturbative contribution should be expressible through characterestic classes and this is why we believe that the symmetry \eqref{symmetry-flat} should hold. 
 
We now consider the anti-instanton Nekrasov partition function. Following \cite{Festuccia:2018rew} we can switch from instantons to anti-instantons by  implementing the following change: $(z_1, z_2) \rightarrow (z_1, \bar{z}_2)$. Therefore we have the following relation between instanton and anti-instanton partition functions
  \bea
   Z^{\rm anti-inst}_{\rm Nekr} (a, \epsilon_1, \epsilon_2, \bar{\Lambda}) = Z^{\rm inst}_{\rm Nekr} (a, \epsilon_1, - \epsilon_2, \bar{\Lambda})~.
  \eea
From which follows
 \bea
   {\cal F}^{\rm anti-inst}_{\rm Nekr} (a, \epsilon_1, \epsilon_2, \bar{\Lambda}) =  - {\cal F}^{\rm inst}_{\rm Nekr} (a, \epsilon_1,
     - \epsilon_2, \bar{\Lambda})~.
 \eea
  Thus for  $ {\cal F}^{\rm inst}_{\rm Nekr} (a, \chi_{\rm eq},  \sigma_{\rm eq} , \Lambda)$ we have
   \bea
    {\cal F}^{\rm anti-inst}_{\rm Nekr} (a, \chi_{\rm eq},  \sigma_{\rm eq} , \bar{\Lambda}) = - {\cal F}^{\rm inst}_{\rm Nekr} (a, - \chi_{\rm eq},  \sigma_{\rm eq} , \bar{\Lambda})~. 
   \eea
 Now comparing to \eqref{sect6-PF-conj-intro}  we can define the following object 
 \bea
  {\cal F} ({\cal A}, \chi_{\rm eq},  \sigma_{\rm eq} , {\mathbf \Lambda}) = \frac{1}{2} \Big (1+ \Omega \Big ) {\cal F}^{\rm inst}_{\rm Nekr} ({\cal A},  \chi_{\rm eq},  \sigma_{\rm eq} , {\mathbf \Lambda} ) + \frac{1}{2} \Big (\Omega-1 \Big ) {\cal F}^{\rm inst}_{\rm Nekr} ({\cal A},  -\chi_{\rm eq},  \sigma_{\rm eq} , {\mathbf \Lambda} ) \label{def-crazy-potential}
 \eea  
  where $\Omega= \Omega (1,-1)$ is defined  in \eqref{def-omega-general} and everything is understood as a class (${\mathbf \Lambda}$ is defined in \eqref{defin-Lambda-class}). We have the following degree allocation
  \bea
  \deg {\cal F} = 4~,~~~~\deg {\cal A} =2~,~~~~\deg {\mathbf \Lambda} =2~,~~~~\deg \chi_{\rm eq} =4~,~~~~\deg \sigma_{\rm eq} =4~.
  \eea 
These degrees constrain the dependence on the various classes. Moreover the dependence on the equivariant characteristic classes of the tangent bundle  $\chi_{\rm eq}$, $\sigma_{\rm eq}$ will come through Taylor expansion  (i.e., only through non-negative powers).  We claim that the localization of the abelian non-linear theory defined by ${\cal F}$ gives the same answer as expected from localization of the non-Abelian theory \eqref{sect6-PF-conj-intro}.  Moreover the answer is compatible with S-duality provided this acts as a Fourier transform.  In the case of the equivariant Donaldson-Witten theory (when all fixed points are either plus or minus) the effective cohomological potential is 
 just ${\cal F}^{\rm inst}_{\rm Nekr} ({\cal A},  \chi_{\rm eq},  \sigma_{\rm eq} , {\mathbf \Lambda} )$. 
   
   Let us illustrate this with examples.  We start with the classical part.  For $\mathbb{C}^2$ we can fix the classical part to be
   \bea
   Z_{\rm cl} = \Lambda^{-\frac{a^2}{\epsilon_1 \epsilon_2}}~,
   \eea
  where we follow the conventions summarrized in \cite{Bershtein:2018zcz}.  Following the definition \eqref{def-crazy-potential} the classical part is given by 
 \bea
  \frac{i}{2\pi} {\cal F}_{\rm cl} = \Omega(1,-1) \Omega(\tau, - \bar{\tau}) {\cal A}^2~.\label{classical-prep}
 \eea
 For $S^4$ with the standard Killing vector we can have essentially two types of supersymmetry (depending on the choice of $s$ and $\tilde{s}$). On  $S^4$ with $(+,+)$ (instantons on both poles) after localizing \eqref{classical-prep} we get  
 \bea
  \frac{\tau}{\epsilon_1\epsilon_2} a^2 + \frac{\tau}{\epsilon_1 (-\epsilon_2)} a^2 =0~.
 \eea
 While on  $S^4$ with $(+,-)$ (Pestun's case \cite{Pestun:2007rz}) after localizing \eqref{classical-prep} we get 
 \bea
   \frac{\tau}{\epsilon_1\epsilon_2} a^2 + \frac{\bar{\tau}}{\epsilon_1 (-\epsilon_2)} a^2 = \frac{(\tau-\bar{\tau})}{\epsilon_1 \epsilon_2} a^2 ~. 
 \eea
 Similarly we can look at ${\mathbb{CP}}^2$ with different assignments of $\pm$ labels at the three fixed points as in \cite{Festuccia:2018rew}. For the case 
 $(+,+,+)$ we get 
\bea
   \frac{\tau}{\epsilon_1\epsilon_2} a^2 +    \frac{\tau}{(\epsilon_2- \epsilon_1)(-\epsilon_1)} a^2 +    \frac{\tau}{(\epsilon_1 - \epsilon_2) (-\epsilon_2)} a^2 =0
\eea
and for the case  $(+,+,-)$
\bea
   \frac{\tau}{\epsilon_1\epsilon_2} a^2 +    \frac{\tau}{(\epsilon_2- \epsilon_1)(-\epsilon_1)} a^2 +    \frac{\bar{\tau}}{(\epsilon_1 - \epsilon_2) (-\epsilon_2)} a^2 =\frac{(\tau - \bar{\tau})}{(\epsilon_2 - \epsilon_1) \epsilon_2} a^2~. 
\eea
Gluing the perturbative parts is subtle due to analytical issues and the problems we mentioned earlier. The instanton (anti-instanton) parts are formally glued in a straightforward fashion.  
 
Let us mention a few related constructions in this context. Our present discussion is somewhat reminiscent of the u-plane description of the Donaldson-Witten theory \cite{Moore:1997pc}. Actually some of the analysis can be performed by a similar gluing of the leading terms of the Nekrasov partition function  \cite{Manschot:2019pog}. 
  We work however in a different setup and our main interest is the equivariant Donaldson-Witten theory and its generalizations. Therefore at the moment we do not see any direct relation between two constructions.   Another related topic is defining supersymmetric partition functions on compact spaces through appropriate gluing of holomorphic blocks.  This was first introduced for 3D theories in \cite{Beem:2012mb} and  the case of 4D/5D theories is reviewed in \cite{Pasquetti:2016dyl} (see also  \cite{Nieri:2013yra, Nieri:2013vba, Nieri:2015yia}). In light of the present discussion we think that the holomorphic blocks, and especially the ambiguities in their definition, should be better understood from the point of view of cohomological field theory.

\section{Summary}\label{section-7}   

The main goal of this paper was to study S-duality for ${\cal N}=2$ supersymmetric linear and non-linear  $U(1)$ theories  on  curved manifolds within the supersymmetry setting that we have introduced previously in \cite{Festuccia:2018rew}. We found that S-duality is compatible with supersymmetry on the curved manifold. We use two languages to discuss supersymmetry, the first uses the original set of physical fields and the second uses cohomological variables. From the geometrical point of view the cohomological variables are more useful and in this language there is a possibility to incorporate the gravitational corrections through the appropriate equivariant characteristic classes for the tangent bundle. 
   These non-linear  ${\cal N}=2$ supersymmetric  $U(1)$ theories can be localized and we can write their partition function as a finite dimensional integral together with a discrete sum. The classical formulations of S-duality as a Legendre transform, however, is not compatible with the localization result, instead we suggest that S-duality should act as a Fourier transform. The Legendre transform is then the leading term in a semi-classical expansion. We have checked that interpreting  S-duality as a Fourier transform is compatible with the localization result and the discrete sum over fluxes plays a crucial role in this construction. 

 In the last section we offered a speculative construction of an effective cohomological prepotential corresponding to a non-linear supersymmetric $U(1)$ theory that gives exactly the same partition function as a supersymmetric non-abelian theory. Our construction is purely formal and it is based on reverse engineering from a known answer. 
   At the moment we do not understand what is the underlying physics behind this object and what are the benefits in its construction.  This prepotential is constructed from the Nekrasov partition function where the parameters are replaced by the appropriate equivariant classes. Some of these classes are the equivariant characteristic classes of the tangent bundle and some are equivariant classes related to the supersymmetry of the theory. We believe that understanding the significance of this object (if any) is paramount. It is also important  to study if  there are any non-trivial constraints on this object that can be derived from first principles.  We also hope that the existence of the cohomological prepotential may bring some light on the original derivation of the localization result since it seems that we still perform some version of the  Nekrasov ADHM calculation albeit with some parameters replaced by equivariant classes. At the moment this is pure speculation and it requires further study.
      
 Another open question is how to use the fact that S-duality is related to the Fourier transform. In flat space S-duality in the form of Legendre transform played a crucial role in the derivation of the explicit form of the Seiberg-Witten prepotential. On curved space with all gravitational corrections turned on we should replace it with a Fourier transform. However a large part of physical intuition available in flat space fails in curved space. For example, we do not know how to discuss the particle spectrum on curved space etc. Thus it would be interesting to understand if we can use the Fourier transform to predict something interesting for  the Nekrasov partition function and for the partition functions on compact spaces. 
   
 In this work we concentrated on the case of a compact manifold. Let us make a few observations about the non-compact case.  On $\mathbb{C}^2$ the Nekrasov partition function satisfies the Nakajima-Yoshioka blowup equation \cite{Nakajima:2003pg, MR2183121, Gottsche:2006bm, Nakajima:2009qjc}. The simplest version of the blowup equation for $\mathbb{C}^2$ has the following form
   \bea
    Z  (a, \epsilon_1, \epsilon_2) = \sum\limits_{n \in \mathbb{Z}} Z (a + n \epsilon_1, \epsilon_1, \epsilon_2-\epsilon_1) 
     Z(a+ n\epsilon_2, \epsilon_1 - \epsilon_2,  \epsilon_2)~,\label{blowup-C}
  \eea
  where $Z$ is full Nekrasov partition function on  $\mathbb{C}^2$. In principle in this formula $a$ and $\epsilon_{1}$, $\epsilon_2$ are complex  numbers. We would like to study how it behaves under Fourier transform. Let us assume for the moment that $a, \epsilon_1, \epsilon_2$ are real (or imaginary). If we plug the formula (\ref{blowup-C}) into (\ref{FT-Z-function}) (see Appendix~\ref{app:Fourier}) and further rewrite everything using Fourier transformed quantities we get the following 
   \bea
    \hat{Z}  (\xi, \epsilon_1, \epsilon_2) =  \sum\limits_{n \in \mathbb{Z}} \hat{Z} (\xi + n \epsilon_1, \epsilon_1, \epsilon_2-\epsilon_1) 
     \hat{Z}(\xi + n\epsilon_2, \epsilon_1 - \epsilon_2,  \epsilon_2)~.\label{FT-blowup-C}
  \eea
 Thus, modulo analytical issues, the blowup equation (\ref{blowup-C}) is invariant under S-duality (the Fourier transform).  
    
   One can establish a similar formal relation for toric non-compact spaces. For example,  consider a theory on $A_1=\mathbb{C}^2/\mathbb{Z}_2$ (or its resolution). The partition function on $A_1$ is obtained gluing two Nekrasov's functions on $\mathbb{C}^2$
  \bea
Z_{A_1} (a, \epsilon_1, \epsilon_2) = \sum\limits_{n \in \mathbb{Z}} Z (a + 2 n \epsilon_1, 2\epsilon_1, \epsilon_2-\epsilon_1) Z(a+ 2n\epsilon_2, \epsilon_1 - \epsilon_2, 2 \epsilon_2)~. 
  \eea   
Now if we perform the Fourier transform of both the left hand side and the right hand side we will arrive to the following relation 
 \bea
 \hat{Z}_{A_1} (a, \epsilon_1, \epsilon_2) = \sum\limits_{n \in \mathbb{Z}} \hat{Z} (a + 2 n \epsilon_1, 2\epsilon_1, \epsilon_2-\epsilon_1) \hat{Z}(a+ 2n\epsilon_2, \epsilon_1 - \epsilon_2, 2 \epsilon_2)~,
\eea   
which we interpret as the action of S-duality. Exactly the same manipulations will work for an $A_n$ singularity. It remains to be seen what is the physical and mathematical meaning of these formal manipulations. 

\bigskip
{\bf Acknowledgements:} We are very grateful to Jian Qiu for his initial participation in the project and for numerous useful discussions. The work of Guido Festuccia is supported by the ERC under the STG grant 639220 and by Vetenskapsr\aa{}det under grant 2018-05572. The work of Maxim Zabzine  is supported by the grant  ``Geometry and Physics"  from the Knut and Alice Wallenberg foundation.

 \appendix
 
\section{Notations for spinors}\label{app:conv}
\label{sec:notations}

We define the Levi--Civita symbol as ~$\epsilon_{1234} = 1$.  Our conventions for Weyl spinors follow those of~\cite{Wess:1992cp}, adapted to Euclidean signature. Left-handed spinors are denoted by undotted indices~$\zeta_\alpha$. Right-handed spinors $\bar \zeta_{\dot\alpha}$ have dotted indices. In Euclidean signature $\zeta$ and~$\bar \zeta$ are not related by complex conjugation. Lower dotted and undotted indices are raised by acting on the left with the tensors $\epsilon_{\alpha \beta} $ and $\epsilon_{\dot \alpha \dot \beta} $ , that are defined as follows $\epsilon^{12} = \epsilon_{21} = \epsilon^{\dot 1 \dot 2} = \epsilon_{\dot 2 \dot 1} = + 1$.   The inner product of~$\zeta$ and~$\eta$ is~$\zeta \eta=\zeta^\alpha \eta_\alpha$. The inner product of~$\bar \zeta$ and~$\bar \eta$ is given by~${\bar \zeta} \bar \eta={\bar \zeta}_{{\dot \alpha}} \bar \eta^{\dot \alpha}$.  The sigma matrices are given by
\begin{equation}
  \label{sigmamat}
  \sigma^\mu_{\alpha \dot{\alpha}} = (\vec{\sigma}, -i 1\!\!1 )~,
  \qquad
  \bar \sigma^{\mu\dot{\alpha} \alpha} = (-\vec{\sigma}, -i 1\!\!1 )~, 
\end{equation}
with ~$\vec{\sigma} = (\sigma^1, \sigma^2, \sigma^3)$ being the Pauli matrices.
We have,
\begin{equation}
  \sigma_\mu\bar \sigma_\nu + \sigma_\nu \bar \sigma_\mu = -2\delta_{\mu\nu}~, \qquad
  \bar \sigma_\mu \sigma_\nu + \bar \sigma_\nu \sigma_\mu = -2\delta_{\mu\nu}~.
\end{equation}
We also define the matrices
\begin{equation}
  \sigma_{\mu\nu} = \frac{1}{4} (\sigma_\mu \bar \sigma_\nu - \sigma_\nu\bar \sigma_\mu)~, \qquad
  \bar \sigma_{\mu\nu} = \frac{1}{4} (\bar \sigma_\mu \sigma_\nu - \bar \sigma_\nu \sigma_\mu)~.
\end{equation}
The are either self-dual or anti self-dual:
\begin{equation}
  \frac{1}{2} \epsilon_{\mu\nu\rho\lambda} \sigma^{\rho \lambda } = \sigma_{\mu\nu}~, \qquad
  \frac{1}{2} \epsilon_{\mu\nu\rho\lambda} \bar{\sigma}^{\rho\lambda} = - \bar \sigma_{\mu\nu}~.
\end{equation}
These matrices can be used to separate a two-form $\omega$  in its $(2,0)$ and $(0,2)$ components
 \begin{equation}
  \omega^{+}_{\alpha \beta} = \frac{1}{2} \omega_{\mu\nu}\sigma^{\mu\nu}_{\alpha \beta}~,\qquad
  \omega^{-}_{{\dot \alpha} {\dot \beta}}= \frac{1}{2} \omega_{\mu\nu}\bar \sigma^{\mu\nu}_{{\dot \alpha} {\dot \beta}}~.
\end{equation}

\section{$\mathcal N=2$ rigid supergravity}
\label{n2sugra}


We consider $\mathcal{N}=2$ theories with a conserved $SU(2)$ R-current. The supergravity to which they can be coupled is described in~\cite{deWit:1979dzm, deWit:1980lyi, deWit:1984rvr, Freedman:2012zz}.  The corresponding rigid supergravity backgrounds have been considered in~\cite{Klare:2013dka, Pestun:2014mja, Butter:2015tra}.
They are specified by a Riemannian manifold equipped with a metric $g$ and a spin structure. Additionally they include an $SU(2)_R$ connection and other auxiliary fields:  a two-form $W_{\mu\nu}$, a scalar $N$, a one-form $G_\mu$, a scalar $S_{ij}$ transforming as an $SU(2)_R$ triplet and finally a closed two-form ${\bf F}_{\mu\nu}$.

A left-handed spinor $\zeta^{i}_{\alpha}$ and a right-handed spinor $\bar{\chi}_{i}^{\dot{\alpha}}$, (here $i$ is an $SU(2)_{R}$ index hence both spinors transform in the fundamental representation of the $SU(2)_{R}$ R-symmetry) parametrize the supergravity variations.  We define spinors $\eta^{i}$ and $\bar{\eta}^{i}$ as follows:
\bea
    &&\eta_i= -S_{i j}\zeta^j+({\bf F}^+ - W^+)\zeta_i- 2 G_\mu\sigma^\mu {\bar \chi}_i~, \\[2pt]
    &&\bar \eta^i=- S^{i j}\bar \chi_j-({\bf F}^- -W^-)\bar \chi^i+2 G_\mu\bar \sigma^\mu {\zeta}^i~.
\eea
Here $W^+={1\over 2} W_{\mu\nu} \sigma^{\mu\nu}$ and $W^-={1\over 2} W_{\mu\nu}\bar \sigma^{\mu\nu}$ (similarly for $\bf F$).

For a rigid supergravity background to admit a supersymmetry the following Killing spinor equations must be satisfied by the variation parameters $\zeta^{i}_{\alpha}$ and  $\bar{\chi}_{i}^{\dot{\alpha}}$. A first set of equations is
\begin{equation}
  \begin{split}
    \label{killinone}
    &(D_\mu -i G_\mu) \zeta_{i}-{i\over 2} W^+_{\mu\rho} \sigma^{\rho}{\bar \chi}_{i} ={i\over 2} \sigma_\mu {\bar \eta}_i~, \\
    &(D_\mu + i G_\mu) {\bar \chi}^{i}+ {i\over 2} W^-_{\mu\rho} {\bar \sigma}^\rho{ \zeta^{i}} ={i\over 2} {\bar \sigma_\mu} {\eta}^i~,
  \end{split}
\end{equation}
where $D_{\mu}$ includes the $SU(2)_{R}$ connection. A second set is
\begin{equation}
  \begin{split}
    \label{killintwo}
    &\Big(N-{1\over 6} R\Big)\bar \chi^i=
    4i \partial_\mu G_\nu \bar \sigma^{\mu\nu} \bar \chi^i
    +{i} \big(\nabla^\mu+2 i G^\mu\big) W^{-}_{\mu\nu} \bar \sigma^\nu\zeta^i
    +i \bar \sigma^\mu\big(D_\mu+{i} G_\mu\big) \eta^i ~,\\
    &\Big(N-{1\over 6} R\Big)\zeta_i
    =-4i \partial_\mu G_\nu \bar \sigma^{\mu\nu} \zeta_i
    -{i} \big(\nabla^\mu
    -2i G^\mu\big) W^{+}_{\mu\nu} \sigma^\nu\bar \chi_i
    +i \sigma^\mu\big(D_\mu-{i} G_\mu\big) \bar \eta_i ~,
  \end{split}
\end{equation}
here $R$ is the Ricci scalar.

Among the bilinears constructed out of the spinors $\zeta^{i}$ and $\bar \chi_{i}$ there are scalars
\begin{equation}
  \label{stsdef}
  s = 2 \zeta^i \zeta_i~,\qquad
  \tilde{s} = 2\bar \chi^i \bar \chi_i~, 
\end{equation}
and the vector field
\begin{equation}    
  v^{\mu} = 2\bar \chi^i\bar\sigma^\mu \zeta_i~. 
\end{equation}
These satisfy $||v||^2=s\tilde s$.  When~\eqref{killinone} are satisfied $v^\mu$ is a Killing vector and $s,~\tilde s$ are constant along $v$.
 
Given a Riemannian spin manifold ${\cal M}$ admitting a Killing vector $v$ with isolated fixed points and functions $s,~\tilde s$ satisfying the constraints above one can construct Killing spinors $\zeta^i$ and $\bar \chi_i$ satisfying~\eqref{killinone} and~\eqref{killintwo}. The details of this construction including the corresponding values for the background supergravity fields are found in~\cite{Festuccia:2018rew}~.

\section{{ ${\cal N}=2$ chiral and vector multiplets}}\label{app:chivec} 

\subsection{Chiral multiplet}

Here we review the ${\cal N}=2$ chiral multiplet of weight $w$~\cite{deRoo:1980mm}.  The bottom component of this multiplet is a complex scalar $X$. At progressively higher levels it comprises a left-handed Weyl fermion $\lambda_i$ in the fundamental of the $R$-symmetry $SU(2)_R$, a self dual two form $B^+$, a complex scalar $D_{i j}$ transforming as a triplet of $SU(2)_R$, a second left handed  Weyl fermion $\psi_i$ in the fundamental of $SU(2)_R$ and, as top component a complex scalar $T$.

Below we will use the following shorthand notations:
\begin{align}
\label{etadef}
&\hat D_\mu X= \big(\partial_\mu-i 2 w  G_\mu\big)X~,\qquad \qquad \qquad \hat D_\mu T=\big(\partial_\mu-2 i (w-2)  G_\mu\big)T~, \notag \\ 
& \hat D_\mu \lambda_i= \big(D_\mu-i (2w-1) G_\mu \big)\lambda_i~,\qquad\; \hat D_\mu \psi_i= \big(D_\mu-i(2w-3)G_\mu \big)\psi_i~,\\
& \hat D_\mu B^+= \big(\nabla_{\mu} -2i(w-1) G_\mu\big)B^+~, \qquad  \hat D_\mu D_{i j}= \big(D_\mu -2i(w-1)G_\mu\big) D_{i j}~. \notag
\end{align}
The susy variations of the components of the chiral multiplet take the form:

\begin{align}
\label{varchiral}
&\delta X=-\zeta_i\lambda^i~, \notag \\ \notag
& \delta\lambda_i =-2 i \big(\hat D_\mu X\big) \sigma^\mu {\bar \chi}_i+2\, B^+\zeta_i+D_{i j} \zeta^j-2 w X\eta_i~,\\ \notag
&\delta {B^+_{\alpha \beta}}= {i\over 2} (\sigma^\mu {\bar\chi_i})_\alpha\big(\hat D_\mu \lambda^i\big)_\beta+{i\over 2} \zeta_{i \alpha }\psi^i_{\beta}+{1\over 2} (1+w) \eta_{i \alpha} \lambda^i_{\beta}+(\alpha\leftrightarrow \beta)~,\\
& \delta D_{i j}=-i \big(\hat D_\mu \lambda_i \big)\sigma^\mu {\bar \chi_j}-i \zeta_i\psi_j+(1-w) \eta_i\lambda_j+(i\leftrightarrow j)~, \\ \notag
& \delta \psi_i=2 \big(\hat D_\mu B^+\big) \sigma^\mu{\bar \chi}_i-\big(\hat D_\mu D_{i j}\big) \sigma^\mu {\bar \chi}^j+2i T \zeta_i+ 2 i (1-w)B^+\eta_i+i (1+w)D_{i j} \eta^j+\\ \notag
&\qquad +2\big(\hat D_\mu X\big) \sigma^\mu W^- {\bar \chi}_i +2 w X\sigma^\mu \big(\nabla_\mu W^-+2 i G_\mu W^- \big){\bar \chi}_i~,\\ \notag
&\delta T={\bar \chi}^i \bar\sigma^\mu\big(\hat D_\mu \psi_i \big)+i w \eta^i \psi_i+i {\bar \chi}^i  W^-\bar \sigma^\mu\big(\hat D_\mu \lambda_i\big)+i (w-1){\bar \chi}^i \big(\nabla_\mu W^-+2 i G_\mu W^- \big)\bar \sigma^\mu \lambda_i~. \notag
\end{align}

Given two chiral multiplets of weights $w_1$ and $w_2$ we can construct a third of weight $w_1+w_2$ by multiplication:
\begin{align}\label{multrules}
& X^{(3)}=X^{(1)} X^{(2)}~,  \notag  \\
& \lambda^{(3)}_i= X^{(1)}\lambda^{(2)}_i +X^{(2)} \lambda^{(1)}_i~,  \notag \\
& B^{+ (3)}_{\alpha \beta}= X^{(1)} B^{+ (2)}_{\alpha \beta}+X^{(2)} B^{+ (1)}_{\alpha \beta}+{1\over 4} \lambda^{i (1)}_\alpha \lambda^{(2)}_{i\beta}+{1\over 4}\lambda^{i (1)}_\beta \lambda^{(2)}_{i\alpha}~, \\
& D^{(3)}_{i j}= X^{(1)} D^{(2)}_{i j}+ X^{(2)} D^{(1)}_{i j} -{1\over 2} \lambda^{(1)}_i \lambda^{(2)}_j -{1\over 2} \lambda^{(1)}_j \lambda^{(2)}_i~, \notag \\
&\psi^{(3)}_i = X^{(1)} \psi^{(2)}_i \!+X^{(2)} \psi^{(1)}_i \!+i  B^{+(1)} \lambda^{(2)}_i \!+i B^{+(2)}\lambda^{(1)}_i \!-{i\over2} D^{(1)}_{i j} \lambda^{(2)j}\! -{i\over2} D^{(2)}_{i j}\lambda^{(1)j}~,\notag  \\
& T^{(3)}= X^{(1)} T^{(2)}\!+ X^{(2)} T^{(1)}\!+{i\over 2} \lambda^{(1)}_i \psi^{(2)i}\!+{i\over 2}\lambda^{(2)}_i \psi^{(1)i}\!+{{B^{+(1)}}_\alpha}^\beta {{B^{+(2)}}_\beta}^\alpha\!+{1\over 4} D^{(1)}_{i j} D^{(2)i j}~. \notag
\end{align}

This can be generalized as follows. Let  $X^a$ be the bottom components of chiral multiplets of weight $w_a$. Consider ${\cal F}(X)$ a holomorphic function  of the $X^a$ of a given weight $w_{\scriptscriptstyle {\cal F}}$ that is: 
\be
\sum_I {\cal F}_a w^a X^a=w_{\scriptscriptstyle {\cal F}} {\cal F}
\ee
We can then build a chiral multiplet with bottom component ${\cal F}(X)$. The components of the multiplet are:
\bea\label{holFmult}
&& X^{({\cal F})}={\cal F}~,\cr
&& \lambda^{({\cal F})}_i= {\cal F}_a \lambda^{a}_i~,\cr
&&B^{+ ({\cal F})}_{\alpha \beta}= {\cal F}_a B^{+ a}_{\alpha \beta}+{1\over 4} {\cal F}_{a b}\lambda^{a i}_\alpha \lambda^{b}_{i\beta}~, \cr
&& D^{({\cal F})}_{i j}= {\cal F}_a D^{a}_{i j}-\ha {\cal F}_{a b} \lambda^{a}_i \lambda^{b}_j~,\cr
&& \psi^{({\cal F})}_{i\alpha} = {\cal F}_a \psi^{a}_{i\alpha} \!+i \,{\cal F}_{a b}{ {B^{+a}}_\alpha}^\beta \lambda^{b}_{\beta i} \!-{i\over2} {\cal F}_{a b} D^{a}_{i j} \lambda^{b j}_\alpha+{i\over 6} {\cal F}_{a b c}\lambda^{a j}_\alpha (\lambda^{b }_j \lambda^{c}_i) ~,\cr
&& T^{({\cal F})}= {\cal F}_a T^{a}+{i\over 2} {\cal F}_{a b} \lambda^{a}_i \psi^{b i}-{1\over 4}{\cal F}_{ab}B^{+a}_{\mu\nu} B^{+b \mu\nu}+{1\over 8} {\cal F}_{ab} D^{a}_{i j} D^{b i j}+\cr
&&+{1\over 8}{\cal F}_{a b c}B^{+ a}_{\mu\nu}\lambda^{b i}\sigma^{\mu\nu}\lambda^{c}_i-{1\over 8} {\cal F}_{a b c}D^{a}_{i j} \lambda^{b i}\lambda^{c j}+{1\over 48} {\cal F}_{a b c d} (\lambda^{a}_i\lambda^{b}_j)(\lambda^{c i}\lambda^{d\, j})~.\qquad
\eea

Starting with a  chiral multiplet of weight $w=2$ the following combination transforms in a total derivative and can be used to construct invariant Lagrangians,
\be
\label{dens}T+{1\over 2}(W^-_{\mu\nu}W^{-\mu\nu})X~.
\ee

Finally irrespective of the number of Killing spinors the following configuration for a chiral multiplet with $w=1$ is invariant under supersymmetry:
\begin{eqnarray}\label{bpschiral} && X=1~,\qquad B^+={\bf F}^+ - W^+~,\qquad D_{i j}= -2S_{i j}~,\cr
&& T=2 i(D^\mu+2i G^\mu) G_\mu+{1\over 2} W^-_{\mu\nu}\left({\bf F}^{-\mu\nu}-W^{-\mu\nu} \right)+\left({1\over 6}R-N\right)
\end{eqnarray}

\subsection{Anti-chiral multiplet}
The bottom component of the anti-chiral multiplet is a complex scalar $\bar X$. At progressively higher levels it comprises a right handed Weyl fermion $\bar \lambda_i$ in the fundamental of $SU(2)_R$, an anti-self dual two form $B^-$, a complex scalar $\bar D_{i j}$ transforming as a triplet of $SU(2)_R$, a second  right handed Weyl fermion $\bar \psi_i$ in the fundamental of $SU(2)_R$ and as top component a complex scalar $\bar T$.

We will use the following shorthand notations:
\begin{align}
\label{etadefb}
&\hat D_\mu \bar X= \big(\partial_\mu+2 i w  G_\mu\big)\bar X~,\qquad \qquad\qquad \hat D_\mu \bar T=\big(\partial_\mu+2 i (w-2)  G_\mu\big)\bar T~,\notag \\
& \hat D_\mu \bar \lambda_i= \big(D_\mu+i (2w-1) G_\mu \big)\bar\lambda_i~,\qquad\; \hat D_\mu \bar\psi_i= \big(D_\mu+i(2w-3)G_\mu \big)\bar\psi_i~,\\
& \hat D_\mu B^-= \big(\nabla_{\mu} +2i(w-1)  G_\mu\big)B^-~, \qquad  \hat D_\mu {\bar D}_{i j}= \big(D_\mu +2 i(w-1)  G_\mu\big) {\bar D}_{i j}~. \notag
\end{align}
The susy variations take the form:
\begin{align}\label{varchiralb} 
&\delta \bar X=\bar\chi^i\bar \lambda_i~,\notag \\
& \delta\bar\lambda^i =2 i \big(\hat D_\mu \bar X\big) \bar\sigma^\mu { \zeta}^i+2\, B^-\bar\chi^i-\b D^{i j} \bar\chi_j+2 w \bar X\bar \eta^i~,\notag \\
&\delta {B^{-\alphadot \betadot}}= {i\over 2} (\bar\sigma^\mu {\zeta^i})^\alphadot\big(\hat D_\mu\bar \lambda_i\big)^\betadot+{i\over 2} \bar\chi^{i \alphadot }\bar\psi_i^{\betadot}+{1\over 2} (1+w) \bar\eta^{i \alphadot} \bar\lambda_i^{\betadot}+(\alphadot\leftrightarrow \betadot)~, \notag \\
& \delta \bar D^{i j}=i \big(\hat D_\mu \bar \lambda^i \big)\bar \sigma^\mu {\zeta^j}+i \bar\chi^i\bar \psi^j-(1-w) \bar\eta^i\bar\lambda^j+(i\leftrightarrow j)~, \\
& \delta \b\psi^i=2 \big(\hat D_\mu B^-\big)\b \sigma^\mu{ \zeta}^i+\big(\hat D_\mu \b D^{i j}\big)\b \sigma^\mu { \zeta}_j-2i \b T \b\chi^i+ 2 i (1-w)B^-\b \eta^i-i (1+w)\b D^{i j} \b \eta_j+\notag \\
&\qquad +2\big(\hat D_\mu \b X\big)\b \sigma^\mu W^+ {\zeta}^i +2 w \b X\b \sigma^\mu \big(\nabla_\mu W^+-2 i G_\mu W^+ \big){\zeta}^i~,\notag \\
&\delta \b T={\zeta}_i \sigma^\mu\big(\hat D_\mu\b \psi^i \big)-i w\b \eta_i \b \psi^i-i {\zeta}_i  W^+ \sigma^\mu\big(\hat D_\mu \b \lambda^i\big)-i (w-1){\zeta}_i \big(\nabla_\mu W^+-2 i G_\mu W^+ \big)\sigma^\mu\b \lambda^i~.\notag
\end{align}

Given two anti-chiral multiplets of weights $w_1$ and $w_2$ we can construct a third of weight $w_1+w_2$ by multiplication (We will only need some components):
\begin{align}
\label{multrulesb}
& \bar X^{(3)}=\bar X^{(1)} \bar X^{(2)}~,\\
& \bar T^{(3)}= \bar X^{(1)} \bar T^{(2)}\!\!+ \bar X^{(2)} \bar T^{(1)}\!\!+{i\over 2} \bar \lambda^{(1)i} \bar \psi^{(2)}_{i}\!\!+{i\over 2}\bar\lambda^{(2)i}\bar \psi^{(1)}_{i}\!\!+{{B^{-(1)}}^\alphadot}_\betadot {{B^{-(2)}}^\betadot}_\alphadot \!+{1\over 4} \bar D^{(1)}_{i j} \bar D^{(2)i j}~. \notag
\end{align}

Starting with an  anti-chiral multiplet of weight $w=2$ the following combination transforms in a total derivative and can be used to construct invariant Lagrangians,
\be\label{densb}
\bar T+{1\over 2}(W^+_{\mu\nu}W^{+\mu\nu})\bar X~.
\ee

\subsection{Vector multiplet}

A vector multiplet can be obtained from one chiral and and one  anti-chiral multiplet both of weight $w=1$  by imposing constraints (in Lorentzian signature the two multiplets are related by complex conjugation hence the constraint is some sort of reality condition). These constraints express the higher components $T, \psi^i$ of the multiplets in terms of lower ones, hence the resulting multiplet is shorter.
\begin{align}
\label{vectorconst}
&\psi^i= -\sigma^\mu ( D_\mu+i G_\mu) \bar \lambda^i ~,\qquad \bar\psi_i= -\bar\sigma^\mu ( D_\mu-i G_\mu) \lambda_i \notag \\[-2pt]
& D^{i j}= \bar D^{i j}~,\notag \\[-2pt]
& T= (D^\mu+2i G^\mu)(\partial_\mu+2 i G_\mu)\b X+{1\over 2} W^-_{\mu\nu}B^{-\mu\nu}+\left({1\over 6}R-N\right) \bar X~,\\[-2pt]
& \bar T= (D^\mu-2i G^\mu)(\partial_\mu-2 i G_\mu)X+{1\over 2} W^+_{\mu\nu}B^{+\mu\nu}+\left({1\over 6}R-N\right) X~,\notag \\[-2pt]
& B_{\mu\nu}+(X W^-_{\mu\nu}+\bar X W^+_{\mu\nu})=dA_{\mu\nu}~. \notag
\end{align}

Here $B=B^++B^-$ and $dA_{\mu\nu}=\partial_\mu A_\nu -\partial_\nu A_\mu$ . 
There is a more general constraint that also results in a vector multiplet. It depends on a constant phase $e^{i\xi}$:
\begin{align}\label{vectorconstb}
&\psi^i= -e^{i\xi} \sigma^\mu ( D_\mu+i G_\mu) \bar \lambda^i ~,\qquad \bar\psi_i= -e^{-i\xi} \bar\sigma^\mu ( D_\mu-i G_\mu) \lambda_i \notag \\[-2pt]
& D^{i j}= e^{i\xi} \bar D^{i j}~,\notag \\[-2pt]
&e^{-i\xi} T=  (D^\mu+2i G^\mu)(\partial_\mu+2 i G_\mu)\bar X+{1\over 2} W^-_{\mu\nu}B^{-\mu\nu}+\left({1\over 6}R-N\right) \bar X~,\\[-2pt]
&e^{i\xi} \b T= (D^\mu-2i G^\mu)(\partial_\mu-2 i G_\mu)X+{1\over 2} W^+_{\mu\nu}B^{+\mu\nu}+\left({1\over 6}R-N\right) X~,\notag \\[-2pt]
& e^{-{i\over 2}\xi} \left( B^+_{\mu\nu}+ \bar X W^+_{\mu\nu} \right)+e^{{i\over 2}\xi} \left(B^-_{\mu\nu}+ X W^-_{\mu\nu}\right)=dA_{\mu\nu}~.\notag
 \end{align}

\section{Cohomohological description of chiral multiplet}\label{app:chiral} 
In order to study the $S$-duality properties of ${\cal N}=2$ abelian gauge theory in curved space, in section~\ref{section-2} we replaced each vector multiplet in the theory by one chiral multiplet and one anti-chiral multiplet (both of weight $1$). As shown in \cite{Festuccia:2018rew} vector multiplets can be recast using cohomological variables in terms of which the properties of supersymmetric observables are more transparent. Here we want to do the same for the combination of chiral and anti-chiral multiplets.

A chiral plus an anti-chiral multiplet comprise the following fields:
\begin{itemize}
\item{ Four complex scalars $X,\b X$ and $T, \b T$ (here and below a bar does not denote complex conjugation),}
\item{ Two $SU(2)_R$ triplet scalars $D_{i j}$ and $\b D_{i j}$,}
\item{ A self-dual two form $B^+$ and an anti self-dual two form $B^-$,}
\item{ Two left handed Weyl fermions $\lambda_i, \psi_i$ and two right handed Weyl fermions $\b\lambda_i, \b \psi_i$ all transforming as doublets of $SU(2)_R$.}
\end{itemize}
There is an invertible map from these fields to the following cohomological multiplets:
\begin{itemize}
\item{ A long multiplet comprising an even scalar $\phi$, an odd one-form $\Psi$, an even two form $F$,  an odd three-form $\rho$ and an even four form $D$. We can arrange these fields in the multi-form $ {\cal A} = \phi + \Psi + F + \rho + D$. Supersymmetry acts on this multi-form as the equivariant differential $d_v=d+\iota_v$
\be
\delta {\cal A}= d_v {\cal A}
\ee}
\item{Two multiplets $(\varphi, \eta)$ and $(\hat \eta, \hat \varphi)$. Here all the components are scalars. The fields $\varphi$ and $\hat \varphi$ are even while $\eta$ and $ \hat \eta$ are odd. Supersymmetry acts as follows:
\be
\delta \varphi=\eta~,\quad \delta \eta = {\cal L}_v\varphi\qquad {\rm and } \qquad  \delta \hat \eta=\hat \varphi~,\quad \delta \hat\varphi = {\cal L}_v\hat \eta
\ee }
\item{Two multiplets $(\chi, H)$ and $(\hat H, \hat \chi)$. Here all the components are two forms in the image of the projector $P^+_\omega$ defined in~\eqref{projector}. The fields $h$ and $\hat h$ are even while $\chi$ and $ \hat \chi$ are odd. Supersymmetry acts as follows:
\be
\delta \chi=H~,\quad \delta H = {\cal L}_v\chi\qquad {\rm and } \qquad  \delta \hat H=\hat \chi~,\quad \delta \hat\chi = {\cal L}_v\hat H
\ee }
\end{itemize}
Note that these cohomological variables are all invariant under $SU(2)_R$. Additionally they are all forms of various degrees and hence can be defined even when the manifold is not spin.

As discussed in Appendix~\ref{app:chivec} one can impose constraints on the combination of one chiral and one anti-chiral multiplet of weight $1$ to obtain a single vector multiplet. At the level of the cohomological variables introduced above these constraints amount to setting to zero the $(\hat \eta,\hat \varphi)$ and $(\hat H, \hat \chi)$ multiplets and also impose the following on the long multiplet components
\be
D=0,\quad \rho=0~,\quad F=dA~.
\ee
Indeed because $dF=0$ the long multiplet is shortened to $(\phi,\Psi, F)$ with $F=dA$. These fields together with the multiplets $(\varphi, \eta)$ and $(\chi, H)$ 
are indeed the cohomological variables introduced for the vector multiplet in~\cite{Festuccia:2018rew}.

The map from the chiral plus anti-chiral components to the cohomological fields can be written explicitly using the Killing spinor $(\zeta_i, \bar \chi_i)$. Below we show what the structure of this map is up to terms proportional to supergravity background fields and curvature. The long multiplet is  
\bea
&&\phi= s \b X +\tilde s X~,\cr
&&\Psi_\mu=\zeta_i\sigma_\mu \b\lambda^i+\b \chi^i \b \sigma_\mu \lambda_i~,\cr
&&F_{\mu\nu}=B^+_{\mu\nu}+B^-_{\mu\nu}+{1\over s+\tilde s} (D_{i j}-\b D_{ij })(\b\chi^i\b\sigma_{\mu\nu}\b\chi^j- \zeta^i \sigma_{\mu\nu}\zeta^j)~,\cr
&&\rho_{\mu\nu\rho}={i{\epsilon_{\mu\nu\rho}}^\lambda\over s+\tilde s} (\zeta_i \sigma_\lambda (\b \psi^i+\b \sigma^\nu D_\nu \lambda^i)+\b \chi^i \b \sigma_\lambda (\psi_i +\sigma^\nu D_\nu \b\lambda_i))~,\cr
&& D_{\mu\nu\rho\lambda}= -i \epsilon_{\mu\nu\rho\lambda}(T-\b T+\nabla^2 (X-\b X))~.
\eea
The other multiplets components instead are
\bea
&&\varphi=-i (X-\b X)~,\cr
&&\eta=\zeta_i \lambda^i +\b \chi^i \b\lambda_i~.
\eea
\bea
&&H_{\mu\nu}=(P^+_\omega)_{\mu\nu}^{\rho\lambda}\left(-F_{\rho\lambda}+{s D_{i j} +\tilde s \b D_{i j} \over s^2+\tilde s^2}(\b\chi^i\b\sigma_{\rho\lambda}\b\chi^j+ \zeta^i \sigma_{\rho\lambda}\zeta^j) +{2 {\epsilon_{\rho\lambda}}^{\alpha\beta}\over s +\tilde s}v_\alpha \partial_\beta \varphi \right)~,\cr
&&\chi_{\mu\nu}= {2\over s^2+\tilde s^2}((s+\tilde s)(\b\chi^i \b \sigma_{\mu\nu} \b\lambda_i-\zeta_i\sigma_{\mu\nu} \lambda^i) +v_\mu \Psi_\nu-v_\nu \Psi_\mu)~.
\eea
\bea
&& \hat\varphi = \epsilon^{\mu\nu\rho\sigma}(i v_\mu D_\nu(\b D_{ij}-D_{i j})(\b\chi^i\b\sigma_{\rho\sigma}\b\chi^j+ \zeta^i \sigma_{\rho\sigma}\zeta^j))+(s T+\tilde s \b T) -\nabla^2(s\b X+\tilde s X)~,\cr
&&\hat\eta=\zeta_i (\psi^i+\sigma^\mu D_\mu \b \lambda^i )-\b\chi_i (\b \psi^i +\b\sigma^\mu D_\mu \lambda^i)~.
\eea
\bea
&& \hat H_{\mu\nu}=(D_{i j}-\b D_{ij })(\b\chi^i\b\sigma_{\mu\nu}\b\chi^j+\zeta^i \sigma_{\mu\nu}\zeta^j)~,\cr
&& \hat\chi_{\mu\nu}=i (s+\tilde s)(\b\chi^i \b\sigma_{\mu\nu}(\b \psi_i +\b\sigma^\rho D_{\rho}\lambda_i)-\zeta_i \sigma_{\mu\nu}(\psi^i+\sigma^\rho D_\rho \b\lambda^i) -i (v_\mu \rho_\nu-v_\nu \rho_\mu) )~.
\eea

\section{Legendre transform} \label{app:Legendre}

In this Appendix we collect various standard formulas for the Legendre transform that we use in the paper. 
Consider a convex function $f(x)$. For a fixed $p$ let $x$ maximize $px + f(x)$ and define 
  \bea
   p = - \frac{d f}{dx}~~~\rightarrow~~~x=g(p)
  \eea
  with the obvious relations
  \bea
   g = (-f')^{-1}~~~~~\leftrightarrow ~~~~f'(x)|_{x=g(p)} = - p~.
  \eea
  The Legendre transform of the function $f$ is defined as 
  \bea
   \hat{f} (p) = p g(p) + f( g(p))~,
  \eea
   and this is an involutive transformation.  We have the following standard relations between the derivatives of a function 
    and its Legendre transform
  \bea
   \frac{d  \hat{f} (p)}{dp} = g(p) + (p + f'(x)|_{x=g(p)}) \frac{d g}{dp} = g(p)~,
  \eea
  and
  \bea
   \frac{ d^2 \hat{f} (p)}{dp^2} = \frac{dg}{dp} = - \frac{1}{f''(x)|_{x=g(p)}}~. 
  \eea 
 
 We are interested in a parametric Legendre transform when the function $f(x,\tau)$ depends on an extra parameter $\tau$
  and we want to perform the Legendre transform on the variable $x$.  
 For a fixed $p$ let $x$ maximize $px + f(x, \tau)$.  We have the relations
 \bea
  p = - \frac{ \partial f}{\partial x} (x, \tau)~~~\rightarrow~~~x=g(p, \tau)
 \eea
     and 
     \bea
      \frac{\partial f}{\partial x} (x,\tau)|_{x=g(p,\tau)}=-p~.
     \eea
     We define the Legendre transform as
 \bea
    \hat{f} (p, \tau) = p g(p, \tau) + f (g(x, \tau), \tau)~. 
 \eea
  Next we want to relate the derivatives of the function and its Legendre transform, including the derivatives with respect to the parameter $\tau$. 
  As before we have the relation
  \bea
   \frac{ \partial^2 \hat{f} (p, \tau)}{\partial p^2} = \frac{\partial g}{\partial p} = - \frac{1}{\partial^2_x f(x, \tau)|_{x=g(p,\tau)}}~. 
  \eea 
   The first derivatives with respect $\tau$ are related as follows
   \bea
  \frac{\partial \hat{f}}{\partial \tau} (p,\tau) = p \frac{\partial g}{\partial \tau} (p, \tau) + 
   \frac{\partial f}{\partial x} (x,\tau)|_{x=g(p,\tau)} \frac{\partial g}{\partial \tau} (p, \tau)  +
     \frac{\partial f}{\partial \tau} (x,\tau)|_{x=g(p,\tau)} =  \frac{\partial f}{\partial \tau} (x,\tau)|_{x=g(p,\tau)} 
 \eea
 and the mixed derivatives are 
 \bea
    \frac{\partial^2 \hat{f}}{\partial \tau\partial p} (p,\tau) =  \frac{\partial^2 f}{\partial \tau\partial x} (x,\tau)|_{x=g(p,\tau)} \frac{\partial g}{\partial p} (p, \tau)  =  - \Big ( \frac{1}{\partial^2_x f(x, \tau)} \frac{\partial^2 f}{\partial \tau\partial x} (x,\tau)\Big )|_{x=g(p,\tau)}~. 
 \eea
 We also have the following relations for the second derivatives with respect to the parameter~$\tau$
 \bea
  \frac{\partial^2 \hat{f}}{\partial \tau^2} (p, \tau) = \frac{\partial^2 f}{\partial x \partial \tau} (x, \tau)|_{x=g(p,\tau)} \frac{\partial g}{\partial \tau} (p, \tau) + \frac{\partial^2 f}{\partial \tau^2} (x, \tau)|_{x=g(p,\tau)}
 \eea
 or alternatively
 \bea
  \frac{\partial^2 \hat{f}}{\partial \tau^2} (p, \tau) = \frac{\partial^2 f}{\partial x \partial \tau} (x, \tau)|_{x=g(p,\tau)} \frac{\partial^2 \hat{f}}{\partial p \partial \tau} (p, \tau) + \frac{\partial^2 f}{\partial \tau^2} (x, \tau)|_{x=g(p,\tau)}~.
 \eea
 All these formulas have simple generalizations in the case of many variables.

\section{Fourier transform} \label{app:Fourier}

In this Appendix we collect the conventions for the Fourier transform used in this paper. Let us introduce the Fourier transform of the function $e^{\frac{2\pi}{\hbar} {\cal F} (x)} :\mathbb{R} \rightarrow 
 \mathbb{C}$  with $\hbar$ being a real number
 \bea
 e^{\frac{2\pi}{\hbar} \hat{\cal F} (\xi)} = {\rm FT} \Big (e^{\frac{2\pi}{\hbar}  {\cal F}(x)} \Big ) = \frac{1}{\sqrt{|\hbar|}} \int\limits_{-\infty}^{\infty} ~e^{\frac{2\pi}{\hbar} [-i x \xi + {\cal F}(x)]}~ dx~,
 \eea 
 and the inverse of the Fourier transform as follows
 \bea
  e^{\frac{2\pi}{\hbar}{\cal F}(x)} =   {\rm FT}^{-1} \Big (e^{\frac{2\pi}{\hbar}  \hat{\cal F}(\xi)} \Big )=\frac{1}{\sqrt{|\hbar|}} \int\limits_{-\infty}^{\infty} ~e^{\frac{2\pi}{\hbar} [i x \xi + \hat{\cal F}(\xi)]}~d\xi~.
 \eea
Here we use the standard representation of the delta function
 \bea
 \frac{1}{|\hbar|} \int\limits_{-\infty}^{\infty} e^{\frac{2\pi }{\hbar} i x\xi} ~d\xi = \delta (x)~.
 \eea
The Fourier transform satisfies the standard properties
  \bea
   {\rm FT}^4 = {\rm Id}~,~~~~~{\rm FT}^2 = {\rm P}~,
 \eea
  where ${\rm P} (f(x)) = f(-x)$ is the parity operator.  
  
Similarly when integrating along the imaginary line we have 
   \bea
 e^{\frac{2\pi}{\hbar} \hat{\cal F} ( i\xi)} = {\rm FT} \Big (e^{\frac{2\pi}{\hbar}  {\cal F}(ix)} \Big ) = \frac{1}{\sqrt{|\hbar|}} \int\limits_{-\infty}^{\infty} ~e^{\frac{2\pi}{\hbar} [-i x \xi + {\cal F}(i x)]}~ dx~,
 \eea 
  and 
   \bea
  e^{\frac{2\pi}{\hbar}{\cal F}(i x)} =   {\rm FT}^{-1} \Big (e^{\frac{2\pi}{\hbar}  \hat{\cal F}(i \xi)} \Big )=\frac{1}{\sqrt{|\hbar|}} \int\limits_{-\infty}^{\infty} ~e^{\frac{2\pi}{\hbar} [i x \xi + \hat{\cal F}(i \xi)]}~d\xi~.
 \eea
   In the context of gauge theories by $\hbar$ we mean the combination $\epsilon_1 \epsilon_2$ and thus we use the following conventions
   \bea
    && Z(a, \epsilon_1, \epsilon_2) = e^{\frac{2\pi}{\epsilon_1\epsilon_2} {\cal F}(a, \epsilon_1, \epsilon_2)} = \frac{1}{\sqrt{|\epsilon_1 \epsilon_2|}} 
     \int\limits_{-\infty}^{\infty} e^{\frac{2\pi}{\epsilon_1 \epsilon_2} [ i a \phi + \hat{\cal F}(\phi, \epsilon_1, \epsilon_2)]}~d\phi \nn\\
     && = \frac{1}{\sqrt{|\epsilon_1 \epsilon_2|}} 
     \int\limits_{-\infty}^{\infty} ~e^{\frac{2\pi i}{\epsilon_1 \epsilon_2}   a \phi } ~\hat{Z} (\phi, \epsilon_1, \epsilon_2)~,
   \eea
    and for its Fourier transform 
    \bea
  &&  \hat{Z}(\xi, \epsilon_1, \epsilon_2) = e^{\frac{2\pi}{\epsilon_1\epsilon_2} \hat{\cal F}(\xi, \epsilon_1, \epsilon_2)}
     = \frac{1}{\sqrt{|\epsilon_1 \epsilon_2|}}    \int\limits_{-\infty}^{\infty} ~e^{\frac{2\pi}{\epsilon_1\epsilon_2} [-i a \xi + {\cal F}(a, \epsilon_1, \epsilon_2)]}~ da \nn \\
     && = \frac{1}{\sqrt{|\epsilon_1 \epsilon_2|}}    \int\limits_{-\infty}^{\infty} ~e^{-\frac{2\pi i}{\epsilon_1\epsilon_2}  a \xi}  ~
     Z(a, \epsilon_1, \epsilon_2) ~da~.\label{FT-Z-function}
   \eea
   Additionally for periodic distributions we use
  \bea
   \frac{1}{|T|} \sum\limits_{n \in \mathbb{Z}} e^{\frac{2\pi}{T} int} = \sum\limits_{k \in \mathbb{Z}} \delta (t - k T)~,
  \eea
 where $T$ can be any real non-zero number.

\providecommand{\href}[2]{#2}\begingroup\raggedright

\bibliographystyle{utphys}
\bibliography{Sduality}{}

\endgroup

\end{document}